\documentclass[aps,prd,twocolumn,superscriptaddress,floatfix,nofootinbib]{revtex4}

\usepackage{color}
\usepackage[dvips]{graphicx}

\newcommand{\lsim}{\mathrel{\mathop{\kern 0pt \rlap
      {\raise.2ex\hbox{$<$}}}\lower.9ex\hbox{\kern-.190em $ \sim$}}}

\newcommand{\gsim}{\mathrel{\mathop{\kern 0pt
      \rlap{\raise.2ex\hbox{$>$}}}\lower.9ex\hbox{\kern-.190em $\sim$}}}

\newcommand{\beq}{\begin{equation}}
\newcommand{\eeq}{\end{equation}}

\newcommand{\be}{\begin{equation}}
\newcommand{\ee}{\end{equation}}
\newcommand{\beqarr}{\begin{eqnarray}}
\newcommand{\eeqarr}{\end{eqnarray}}

\newcommand{\widesigmav}{\widetilde{\langle \sigma_{\rm ann} v \rangle}}
\newcommand{\sigmavint}{\langle \sigma_{\rm ann} v \rangle_{\rm int}}

\begin{document}

\title{Discussing direct search of dark matter particles in the Minimal
Supersymmetric extension of the Standard Model with light neutralinos}

\thanks{Preprint number: DFTT 19/2010}


%

%
\author{N. Fornengo}
\affiliation{Dipartimento di Fisica Teorica, Universit\`a di Torino \\
Istituto Nazionale di Fisica Nucleare, Sezione di Torino \\
via P. Giuria 1, I--10125 Torino, Italy}
\author{S. Scopel}
\affiliation{Department of Physics, Sogang University\\
Seoul, Korea, 121-742}
\author{A. Bottino}
\affiliation{Dipartimento di Fisica Teorica, Universit\`a di Torino \\
Istituto Nazionale di Fisica Nucleare, Sezione di Torino \\
via P. Giuria 1, I--10125 Torino, Italy}

\date{\today}

\begin{abstract}

We examine the status of light neutralinos in an effective
Minimal Supersymmetric extension of the Standard Model (MSSM) at the electroweak scale
which was considered in the past and discussed in terms of the available data
of direct searches for dark matter (DM) particles.
Our reanalysis is prompted by new measurements at the Tevatron and B-factories
which might  potentially provide significant constraints on the MSSM model. Here we examine in detail
all these new data and show that the present published results from the Tevatron and B-factories  have only
a mild effect on the original light neutralino population. This population,
which fits quite well the DAMA/LIBRA annual modulation data, would also agree with the preliminary
results of CDMS, CoGeNT and CRESST, should these data, which are at present only hints or excesses of events over the expected backgrounds,  be interpreted as authentic signals of DM.
For the neutralino mass we find a lower bound of 7-8 GeV.
Our results differ from some recent conclusions by other authors because of a few crucial
points which we try to single out and elucidate.

\end{abstract}


\maketitle

\section{Introduction}
\label{sec:intro}

Much interest has recently been raised by some new hints of possible
signals of dark matter (DM) particles in experiments of direct detection
(CDMS \cite{cdms}, CoGeNT  \cite{cogent}, CRESST \cite{cresst}) which previously reported upper bounds only.
These hints are in fact only constituted by excesses of events over what
would be expected from backgrounds. What is intriguing is that these events, if
actually due to DM particles with a coherent interaction with the atomic
nuclei of the detector material, would be concentrated in a physical region
which, expressed in terms of the WIMP mass and of the WIMP-nucleon
elastic cross--section, agrees with the physical region established with
a high statistical significance by the DAMA Collaboration
from a measurement of annual modulation over 13 yearly cycles with the DAMA/NaI and the DAMA/LIBRA experiments \cite{dama2010}.

These results have prompted a large number of phenomenological papers focussed on
WIMPs with a light mass (around 10 GeV) and a WIMP-nucleon elastic cross--section
of order ($10^{-40} -  10^{-41}$) cm$^2$,  whereas previous theoretical and
experimental considerations were prevalently directed toward physical regions
with much higher masses and lower cross--sections.  Turning to a specific candidate,
it has now become common to consider  neutralinos of light mass ($\sim$ 10 GeV).

Actually, already long ago in Ref. \cite{lowneu} it was stressed that, in case of R-parity
conservation, a light neutralino ({\it i.e.} a neutralino with
$m_{\chi} \lsim$ 50 GeV), when it happens to be the Lightest
Supersymmetric Particle (LSP), constitutes an extremely interesting
candidate for the dark matter in the Universe, with direct
detection rates accessible to experiments of present generation. In
Ref. \cite{lowneu} it was also derived a lower bound of $m_{\chi} \sim$ 7
GeV from the cosmological upper limit on the cold dark matter density.
The theoretical framework, considered in Ref. \cite{lowneu}, which allows neutralinos with a mass in the
range 7 GeV $\lsim m_{\chi} \lsim$ 50 GeV is
an effective Minimal Supersymmetric extension of the
Standard Model (MSSM) at the electroweak (EW) scale, where the
usual hypothesis of gaugino-mass unification at the scale of Grand Unification (GUT)
of the SUGRA models, is removed; this effective MSSM is very
manageable, since expressible in terms of a limited number of
independent parameters. This model is the theoretical basis we also adopt for the phenomenological investigations
presented in this work (for simplicity, the model which entails low-mass
($m_{\chi} \lsim$ 50 GeV) neutralino configurations within
this  effective MSSM will hereby be dubbed Light Neutralino Model (LNM));
its main features are briefly summarized in Sect. \ref{sec:model}.

When the DAMA Collaboration published their experimental results
collected with a NaI detector of 100 kg over 7 annual cycles
\cite{dama2004}, in Ref. \cite{bdfs2004} it was proved that indeed the
population of light neutralinos \cite{lowneu} fitted well these data.
The possible interpretation of the annual modulation results in terms
of light neutralinos was further confirmed in Refs.~\cite{zoom,inter},
when the channeling effect was taken into account in the experimental
analysis \cite{dama2008/1} and the first DAMA/LIBRA combined data were
presented \cite{dama2008/2}.

We recall that the present collection of
data by the same Collaboration \cite{dama2010} amounts to an exposure
of 1.17 ton x year, with an evidence for an annual modulation effect
at 8.9 $\sigma$ C.L..  This extended collection of data
entails that the DAMA/LIBRA annual modulation regions in the
plane WIMP mass -- WIMP--nucleon scattering cross--section, reported in the figures of 
Ref. \cite{inter}
with a comparison with the theoretical predictions within LNM,
  essentially maintain their shapes, but with a  statistical significance increased from 6.5 $\sigma$ to 7.5 $\sigma$
  (see Ref. \cite{inter} for a detailed definition of these regions).

As mentioned above, other experimental collaborations have recently
reported some events
that might be in excess of the expected backgrounds. Two candidate
events for dark matter, which would survive after application of
various discrimination and subtraction procedures, were reported by the
CDMS Collaboration \cite{cdms}. In Ref. \cite{noiCDMS} it is shown
that, should these events be actually due to dark matter, they would
be compatible with light neutralinos and the DAMA results.  Likewise,
compatible with the LNM and the previously quoted experimental data,
would be the excess of bulk-like events reported by the CoGeNT
Collaboration \cite{cogent}, again in case these might actually be
significant of a DM effect. Most recently, also the CRESST
Collaboration has presented results which, if interpreted as due to DM
particles, would point to WIMPs with a mass $\lsim$ 15 GeV and a
scalar WIMP--nucleon cross section in the ballpark of the previous
experiments \cite{cresst}.

As previously mentioned,  the new experimental results
of Refs.~\cite{cdms,cogent,cresst}, combined with the DAMA data, have
recently triggered a large number of phenomenological considerations
on coherently-interacting WIMPs of light mass (see for instance
Refs.~\cite{kopp,fitz,nath,kuflik,barger,savage,collarhooper,das,belanger,belikov,gunion,liu}),
with emphasis for the mass range 7--10 GeV. We wish to stress that, though
this is a very interesting mass interval, already pointed out in
Refs.~\cite{lowneu,bdfs2004,zoom,inter}, the physical region compatible with the DAMA results
alone and accessible, in particular, to relic neutralinos is much wider; its
specific size depends on a number of astrophysical features ({e.g.} WIMP
distribution function in the galactic halo) and on the detector response
(role of the channeling effect in detectors which use crystals \cite{chann}).
For instance, it was shown in Ref. \cite{inter}
that for the case of a WIMP halo distribution given by a cored--isothermal
sphere the extended mass range is 7 GeV $\lsim m_{\chi} \lsim$ 60 GeV.
In view of the high significance of the annual modulation  results versus
the still preliminary character of the hints of experiments in
Refs.~\cite{cdms,cogent,cresst},
we will consider in the present paper the status of relic neutralinos in the
whole WIMP light--mass range  7 GeV $\lsim m_{\chi} \lsim$ 50 GeV, as we did
in our previous publications on light neutralinos.

Our present investigation is mainly focussed on the role that recent measurements at
the Tevatron and at the B--factories  BaBar and Belle can have in providing
 constraints on supersymmetric models  more stringent than
 the ones previously considered in Ref. \cite{inter}.   Specifically, we consider the
new data concerning the following processes:
a) the decay $B_s \rightarrow \mu^{-} + \mu^{+}$, the top--to--bottom quark decay with
emission of a charged Higgs ($t \rightarrow b + H^+$), and the searches for
neutral Higgs bosons into a tau-lepton pair, at the Tevatron; b) the rare decays
$B \rightarrow \tau + \nu_{\tau}$ and $B \rightarrow D+ \tau + \nu_{\tau}$
(and $B \rightarrow D+ l + \nu_{l}$, where $l = e,\mu$), at the  B-factories  BaBar and Belle.

These measurements have potentially
a significant role in constraining the supersymmetric
parameter space in the region of high $\tan \beta$ and of light Higgs masses and
consequently in determining the allowed ranges for a number of crucial quantities:
neutralino relic abundance, lower bound on the neutralino mass and elastic
neutralino--nucleon cross--section.
Here the nature of these constraints is critically analyzed,
taking also into account the fact that
many of them are still affected by sizable uncertainties. Their impact in constraining
the parameter space of our supersymmetric model is investigated in a twofold way:
by analytic investigations and by detailed numerical analyses.
We believe that the  analytic derivations are necessary: i) to clarify how the
relevant physical quantities
mentioned above depend on the parameters of the LNM, ii) to establish the impact that
each specific constraint has on these model parameters and then consequently on the
physical quantities,  iii) to direct the numerical analyses to the regions of the
parameter space which are of most relevance, iv) to interpret
the outcomes of the numerical evaluations correctly and, finally, v) to serve as a guide for
an educated guess about how the present situation could evolve, as new experimental
limits on supersymmetric parameters from accelerator
experiments and other precision measurements might become available in the future.

In the course of our discussion
we will also comment on how our results compare with some of the outcomes of other recent papers where
numerical analyses of the previously mentioned constraints have been discussed (see, for instance,
Refs.~\cite{nath,kuflik,belanger,belikov,gunion,albornoz}).

The scheme of our paper is the following.
In Sect. \ref{sec:model} the main features of our effective supersymmetric model are presented.
In Sect. \ref{sec:cosmo} we derive from the analytic expression of the neutralino relic abundance
the lower bound for the neutralino mass in a form which displays its dependence on the main
model parameters. Then in Sect. \ref{sec:cross} we derive an approximate expression for the neutralino--nucleon
cross--section, which provides an easy estimate for this quantity. In Sect. \ref{sec:constraints} we give an overview
of the most relevant constraints which can have an impact on the ranges of our model parameters.
Our evaluations are compared to the current  results from experiments of direct DM particles
in Sect. \ref{sec:experiments}. We draw our conclusions in Sect.\ref{sec:conclusions}.

\section{Light neutralinos in an effective MSSM (LNM)}
\label{sec:model}

%
\begin{figure}[t]
\includegraphics[width=1.1\columnwidth,clip=true,bb=18 60 592 520]{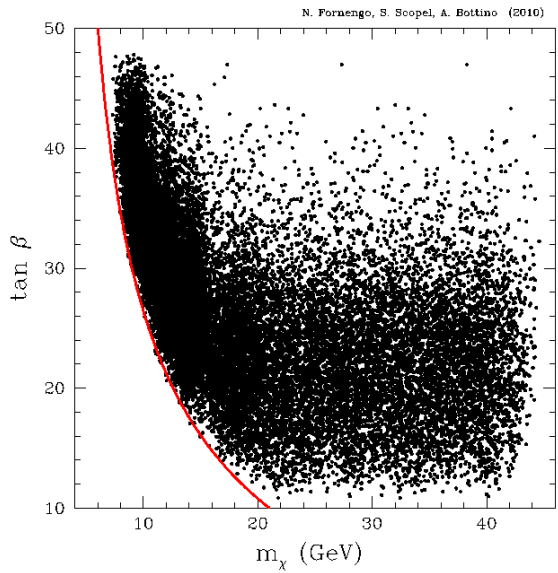}
\caption{Scatter plot of the light neutralino population for the LNM--$\cal A$ scan,
shown in the plane $m_{\chi}-\tan \beta$. The (red) solid line represents the analytic bound from the neutralino relic abundance given in Eq. (\ref{ma}).}
\label{fig:scat_mchi_tanb}
\end{figure}
\begin{figure}[t]
\includegraphics[width=1.1\columnwidth,clip=true,bb=18 60 592 520]{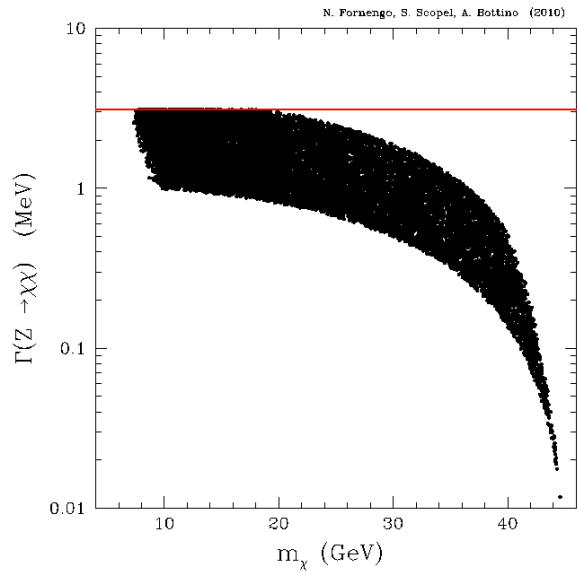}
\caption{Scatter plot for
$\Gamma(Z \rightarrow \chi \chi)$ versus $m_{\chi}$
for the LNM--$\cal A$ scan. The (red) solid horizontal line denotes the
present experimental upper bound to the invisible width of the Z--decay
into non Standard Model particles.}
\label{fig:scat_mchi_gammaz}
\end{figure}
%
%
\begin{figure}[t]
\includegraphics[width=1.1\columnwidth,clip=true,bb=18 60 592 520]{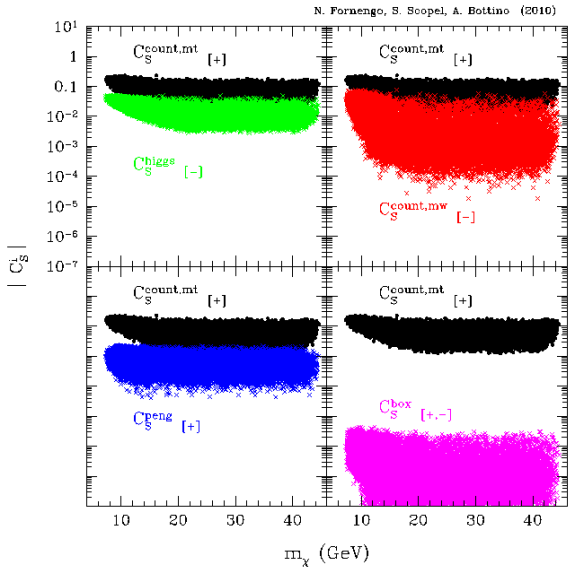}
\caption{Scatter plot of the absolute value of the Wilson coefficients $|C_S^i|$ for each SUSY contribution
to the $BR(B_s \rightarrow \mu^+ \mu^-)$ (colored points), compared to the full calculation 
of the dominant term (black points) (for expressions of these quantities see Refs. \cite{bobeth,arnowitt,buras,eos}), for the LNM--$\cal A$ scan. The sign of each term is indicated in parenthesis as ``$[+]$'' and ``$[-]$''. The values of $|C_S^i|$ are plotted as functions of the neutralino mass $m_\chi$.}
\label{fig:scat_mchi_bsmumu_contrib0}
\end{figure}
\begin{figure}[t]
\includegraphics[width=1.1\columnwidth,clip=true,bb=18 60 592 520]{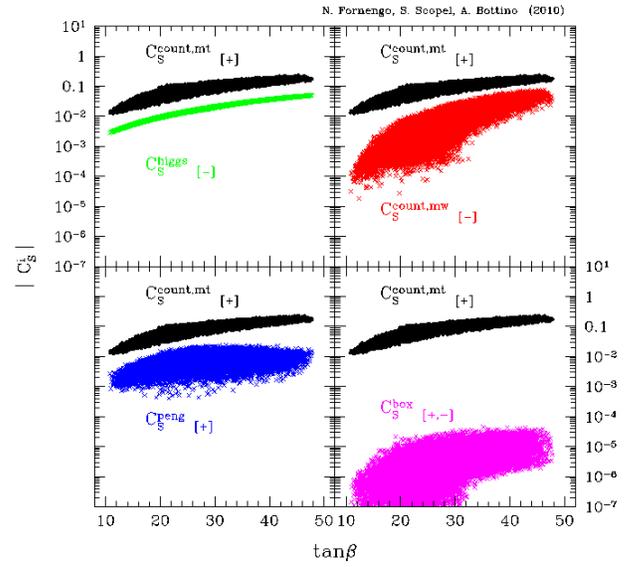}
\caption{The same as in Fig. \ref{fig:scat_mchi_bsmumu_contrib0}, except that the
Wilson coefficients $|C_S^i|$ are plotted as functions of $\tan\beta$.}
\label{fig:scat_tanb_bsmumu_contrib0}
\end{figure}
\begin{figure}[t]
\includegraphics[width=1.1\columnwidth,clip=true,bb=18 60 592 520]{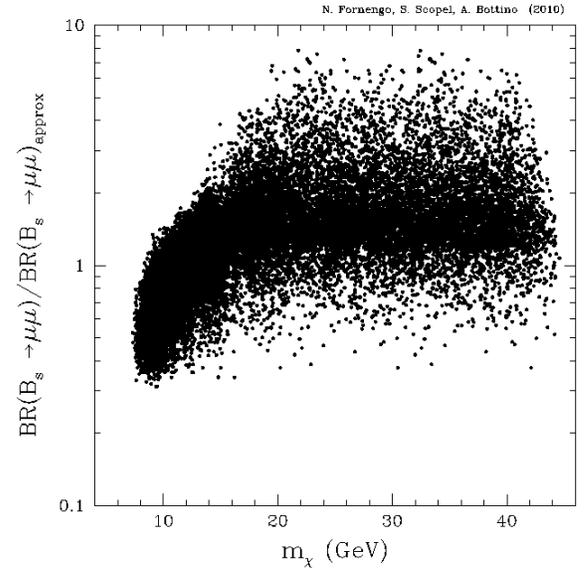}
\caption{Scatter plot of the ratio between the full numerical calculation of 
$BR(B_s \rightarrow \mu^+ \mu^-)$ and the approximate expression of the
dominant term $BR^{(6)}(B_s \rightarrow \mu^+ \mu^-)$ given in Eq. (\ref{mumu4}), 
for the LNM--$\cal A$ scan. The scatter plot is shown
as a function of the neutralino mass $m_\chi$.}
\label{fig:scat_mchi_bsmumu_contrib}
\end{figure}
\begin{figure}[t]
\includegraphics[width=1.1\columnwidth,clip=true,bb=18 60 592 520]{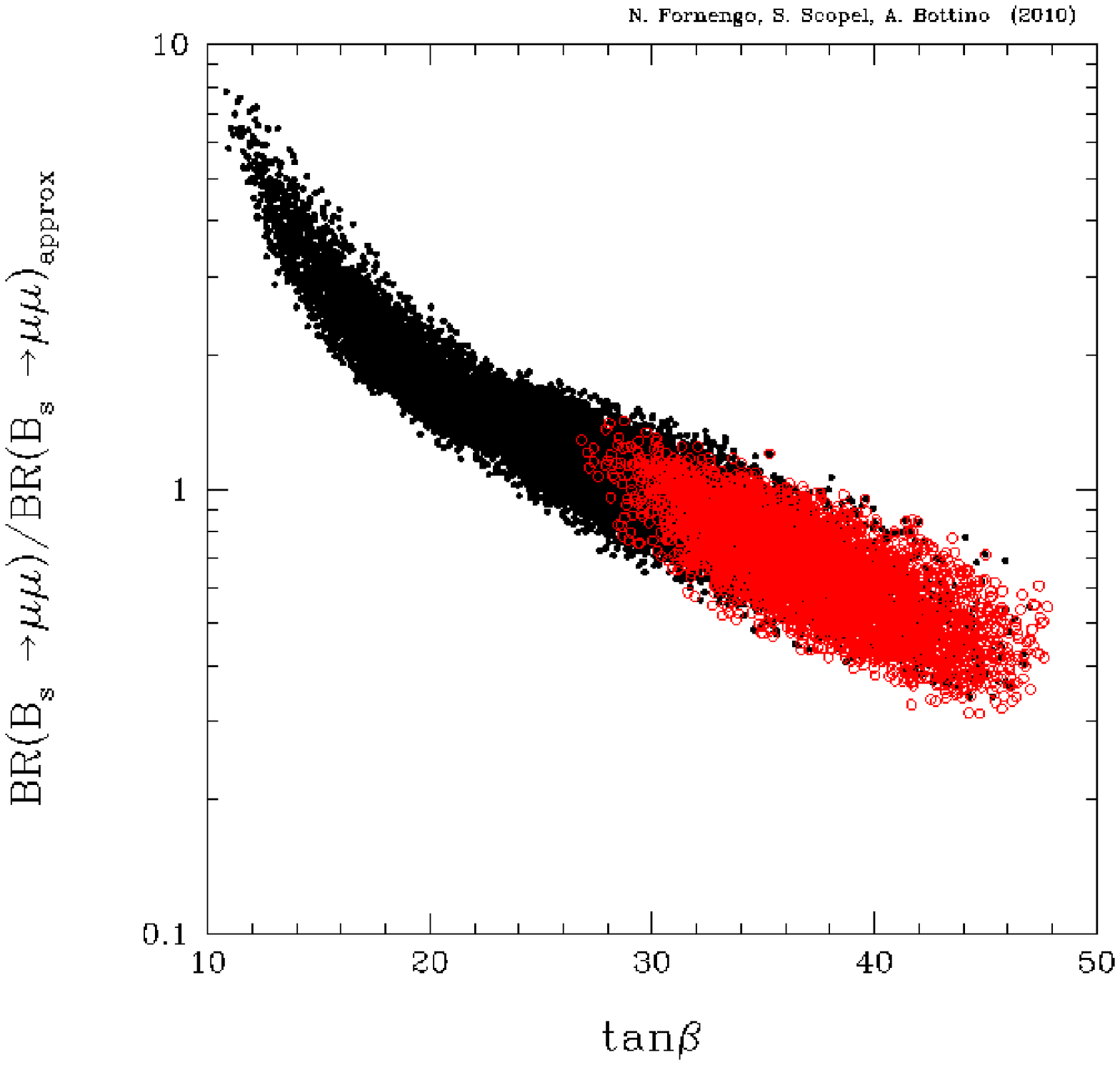}
\caption{The same as in Fig. \ref{fig:scat_mchi_bsmumu_contrib}, except that the
scatter plot is shown as a function of $\tan\beta$.}
\label{fig:scat_tanb_bsmumu_contrib}
\end{figure}
%

The supersymmetric scheme we employ in the present paper is the one
described in Ref. \cite{lowneu}: an effective MSSM scheme (effMSSM) at
the electroweak scale, with the following independent parameters:
$M_1, M_2, M_3, \mu, \tan\beta, m_A, m_{\tilde q}, m_{\tilde l}$ and
$A$. We stress that the parameters are defined at the EW scale.
Notations are as follows: $M_1$, $M_2$ and $M_3$ are the U(1),
SU(2) and SU(3) gaugino masses (these parameters are taken here to be
positive), $\mu$ is the Higgs mixing mass parameter, $\tan\beta$ the
ratio of the two Higgs v.e.v.'s, $m_A$ the mass of the CP-odd neutral
Higgs boson, $m_{\tilde q}$ is a squark soft--mass common to all
squarks, $m_{\tilde l}$ is a slepton soft--mass common to all
sleptons, and $A$ is a common dimensionless trilinear parameter for
the third family, $A_{\tilde b} = A_{\tilde t} \equiv A m_{\tilde q}$
and $A_{\tilde \tau} \equiv A m_{\tilde l}$ (the trilinear parameters
for the other families being set equal to zero).  In our model, no
gaugino mass unification at a Grand Unified scale is assumed.

The following experimental constraints are imposed: accelerators data
on supersymmetric and Higgs boson searches at the CERN $e^+ e^-$ collider
LEP2 \cite{LEPb}; the upper bound on the invisible width for the decay
of the $Z$--boson into non Standard Model particles:
$\Gamma(Z \rightarrow \chi \chi) <$ 3 MeV
\cite{aleph05,pdg} (the role of this bound will be discussed in
Sect. \ref{sec:scenarioA});
measurements of the $b \rightarrow s + \gamma$ decay
process \cite{bsgamma}: 2.89 $\leq BR(b \rightarrow s \gamma) \cdot
10^{4} \leq$ 4.21 is employed here (this interval is larger by 25\%
with respect to the experimental determination \cite{bsgamma} in order
to take into account theoretical uncertainties in the supersymmetric
(SUSY) contributions \cite{bsgamma_theorySUSY} to the branching ratio
of the process (for the Standard Model calculation, we employ the
 NNLO results from Ref. \cite{bsgamma_theorySM})); the measurements of the muon anomalous
magnetic moment $a_\mu \equiv (g_{\mu} - 2)/2$: for the deviation,
$\Delta a_{\mu} \equiv a_{\mu}^{\rm exp} - a_{\mu}^{\rm the}$, of the experimental world average from the
theoretical evaluation within the Standard Model we use here the (2 $\sigma$) range
$31 \leq \Delta a_{\mu} \cdot 10^{11} \leq 479 $, derived from the
latest experimental \cite{bennet} and theoretical \cite{davier} data (the supersymmetric contributions
to the muon anomalous magnetic moment within the MSSM are evaluated here by using the formulae
in Ref. \cite{moroi}; the constraints on the SUSY parameters obtained from the searches for the neutral Higgs--boson at the 
Tevatron \cite{abazov_tautau_2008, abazov_tautau_2009, aaltonen_tautau_2009}; the
upper bound (at $95 \%$ C.L.) on the branching ratio for the decay $B_s \rightarrow {\mu}^{+} + {\mu}^{-}$:
$BR(B_s \rightarrow {\mu}^{+} {\mu}^{-}) < 5.8 \times 10^{-8}$ \cite{cdf_mumu} and the constraints related to
$\Delta M_{B,s} \equiv M_{B_s} - M_{\bar{B}_s}$ \cite{buras_delta,isidori}.
The role of these two  last categories of bounds in constraining the LNM is elucidated in
Sect. \ref{bsdecay}, while the constraints from the searches for neutral Higgs--bosons
at the Tevatron are discusses in Sect. \ref{searchneutralH}.
 The cosmological upper bound on Cold Dark Matter (CDM), which is also implemented in our calculations, is discussed
in Sect. \ref{sec:cosmo}. Other possible constraints  from the Tevatron and B-factories are analyzed in detail in
Sect. \ref{sec:constraints}.

The linear superposition of bino $\tilde B$, wino $\tilde W^{(3)}$
and of the two Higgsino states $\tilde H_1^{\circ}$, $\tilde
H_2^{\circ}$ which defines the neutralino state of lowest mass $m_{\chi}$ is written here as:
\begin{equation}
\chi \equiv a_1 \tilde B + a_2 \tilde W^{(3)} +
a_3 \tilde H_1^{\circ} + a_4  \tilde H_2^{\circ}.
\label{neutralino}
\end{equation}
The properties of these states have been investigated in detail,
analytically and numerically, in Ref. \cite{lhc1} for the case when
the smallest mass eigenstate $\chi_1$ (or $\chi$ in short) is light,
{\it i.e.} $m_{\chi} \equiv m_{\chi_1}\lsim 50$ GeV.  Of that analysis
we report here only the main points which are relevant for the present
paper.

We first notice that the lowest value for $m_{\chi}$ occurs when:
\begin{equation}
m_{\chi} \simeq M_1 \ll |\mu|\, ,\, M_2\, ,
\label{approx}
\end{equation}
since the LEP lower limit on the chargino mass ($m_{{\chi}^{\pm}}
\gsim$ 100 GeV) sets a lower bound on both $|\mu|$ and $M_2$: $|\mu|,
M_2 \gsim$ 100 GeV, whereas $M_1$ is unbound.  Thus, $\chi$
is mainly a Bino; its mixings with the other interaction
eigenstates are given by:
\begin{eqnarray}
\frac{a_2}{a_1} &\simeq& \frac{\xi_1}{M_2} \cot \theta_W,
\label{approx1} \\
\frac{a_3}{a_1} &\simeq& \sin  \theta_W \sin \beta \; \frac{M_Z}{\mu},
\label{approx2} \\
\frac{a_3}{a_4} &\simeq&  - \frac{\mu \sin \beta}{M_1 \sin \beta + \mu \cos \beta},
\label{approx3}
\end{eqnarray}
where $\xi_1 \equiv m_\chi - M_1$ and $\theta_W$ is the Weinberg angle. These expressions readily follow from
the general analytical formulae given in Ref. \cite{lhc1} by taking
$\tan \beta \geq$ 10, as consistent with the scenarios discussed
below.

From the above expressions the following relevant property holds:
$\chi$ is mainly a Bino whose mixing with $\tilde H_1^{\circ}$ is
non--negligible at small $\mu$. In fact, for the ratio $|a_3|/|a_1|$
one has:
\begin{equation}
\frac{|a_3|}{|a_1|} \simeq \sin \theta_W \; \sin \beta \;
\frac{M_Z}{|\mu|} \lsim 0.43 \; \sin \beta,
\label{ratio}
\end{equation}
where in the last step we have taken into account the experimental
lower bound $|\mu| \gsim $ 100 GeV.

It is also useful to explicit the connection between the
neutralino mass $m_{\chi}$ and the parameter $M_1$ at small $m_{\chi}$. From the diagonalization of the neutralino
mass matrix one finds:
\begin{eqnarray}
m_{\chi} = M_1 - \sin \theta_W \; M_Z \; \left(\frac{a_3}{a_1} \; 
\cos \beta   - \frac{a_4}{a_1} \; \sin \beta\right).
\label{mchim1}
\end{eqnarray}
Employing Eqs. (\ref{approx1}) -- (\ref{approx3}) we obtain:
\begin{eqnarray}
m_{\chi} &\simeq& M_1 - \sin^2 \theta_W \; \frac{m^2_Z}{\mu^2} \; 
\left (\frac{2 \; \mu}{\tan \beta} + M_1 \right) \\
&\simeq& M_1 \; (1 - 0.16 \; \mu^{-2}_{100}) - 1.1 \; \mu_{100} \; \left(\frac{35}{\tan \beta}\right) \; {\rm GeV} \nonumber,
\label{mchim12}
\end{eqnarray}
or:
\begin{eqnarray}
 M_1 \simeq \frac{1}{(1 - 0.16 \; \mu^{-2}_{100})}
\left[m_{\chi} + 1.1 \left(\frac{35}{\tan \beta}\right) \mu_{100} \; {\rm GeV} \right],
\label{mchim13}
\end{eqnarray}
where $\mu_{100}$ denotes $\mu$ in units of 100 GeV.

\section{Cosmological bound and lower limit to the neutralino mass}
\label{sec:cosmo}

The neutralino relic abundance is given by:
\begin{equation}
\Omega_{\chi} h^2 = \frac{x_f}{{g_{\star}(x_f)}^{1/2}} \frac{9.9 \cdot
10^{-28} \; {\rm cm}^3 {\rm s}^{-1}}{\widesigmav},
\label{omega}
\end{equation}

\noindent
where $\widesigmav \equiv x_f \sigmavint$, $\sigmavint$ being the integral from the present temperature up to
the freeze-out temperature $T_f$ of the thermally averaged product of
the annihilation cross--section times the relative velocity of a pair
of neutralinos, $x_f$ is defined as $x_f \equiv m_{\chi}/T_f$ and
${g_{\star}(x_f)}$ denotes the relativistic degrees of freedom of the
thermodynamic
bath at $x_f$.  For $\widesigmav$
we will use the standard expansion in S and P waves:
$\widesigmav \simeq \tilde{a} + \tilde{b}/(2 x_f) $.
Notice that in the LNM no coannihilation effects are present in the
calculation of the relic abundance, due to the large mass splitting
between the mass of the neutralino ($m_{\chi}<50$ GeV) and those of
sfermions and charginos.

In our numerical evaluations all relevant contributions to the pair annihilation
cross--section in the denominator of Eq. (\ref{omega}) are included.
However, approximate expressions for $\Omega_{\chi} h^2$ can be derived analytically;
these will prove to be very useful to obtain analytic formulae for
the lower bound for the neutralino mass.

\subsection{ Scenario {$\mathcal{A}$}}
\label{sec:scenarioA}

We first analyze the case of small values for the mass of the CP--odd neutral
Higgs boson $m_A$: 90 GeV $\leq m_A \leq$ 150 GeV. In this case  the main contribution
to  $\sigmavint$ is provided by the $A$--exchange in
the $s$ channel of the
annihilation process $\chi + \chi \rightarrow \bar{b}+b$
(one easily verifies that when $m_{\chi} <
m_b$, $\sigmavint$
entails a relic abundance exceeding the cosmological bound).
Thus one obtains \cite{lowneu}:
\begin{widetext}
\begin{eqnarray}
\Omega_{\chi} h^2 &\simeq& \frac{4.8 \cdot 10^{-6}}{\rm GeV^{2}} \frac{x_f}{{g_{\star}(x_f)}^{1/2}}
\frac{1}{ a_1^2 a_3^2 \tan^2\beta }
m_A^4
\frac{[1-(2m_{\chi})^2/m_A^2]^2}{m_{\chi}^2~[1-m_b^2/m_\chi^2]^{1/2}}
\frac{1}{(1 + \epsilon_b)^2},
\label{eq:omega0}
\end{eqnarray}
\end{widetext}
where $\epsilon_b$ is a quantity which enters in the relationship between the b--quark
running mass and the corresponding Yukawa coupling (see Ref. \cite{higgs} and references quoted therein). In deriving this expression, one has taken into account that here
the following hierarchy holds for the coefficients $a_i$ of $\chi$:
\begin{equation}
|a_1| > |a_3| \gg |a_2|, |a_4|,
\label{hierarchy1}
\end{equation}
as easily derivable from Eqs. (\ref{approx1}--\ref{approx3}).

As far as the value
of ${g_{\star}(x_f)}^{1/2}$ is concerned, we notice that for
light neutralinos  $x_f \simeq 21-22$, so that neutralinos with masses
$m_{\chi} \simeq $ 6 -- 7 GeV have
a freeze--out temperature $T_f \sim T_{QCD}$, where $T_{QCD}$ is the
hadron--quark transition temperature of order 100 -- 300 MeV.  For
definiteness, we describe here the hadron-quark transition by a step
function: if $T_{QCD}$ is set equal to 300 MeV, then for $m_{\chi}
\lsim $ 6 GeV one has ${g_{\star}(x_f)}^{1/2} \simeq 4$, while for
heavier neutralinos ${g_{\star}(x_f)}^{1/2} \simeq 8-9$. In the approximate
analytic expressions discussed hereafter, we set $x_f/g_{\star}(x_f)^{1/2}$ = 21/8
(while in the numerical analysis the actual values obtained after solving the
Boltzmann equation are used).

In selecting the physical parameter space for relic neutralinos, a first fundamental
constraint to be applied is that the neutralino relic abundance does not exceed
the observed upper bound for cold dark matter (CDM), {\it i.e.}
$\Omega_{\chi} h^2 \leq (\Omega_{CDM} h^2)_{\rm max}$. If we apply
this requirement, by using Eq. (\ref{eq:omega0}) we obtain  the following
lower bound on the  neutralino mass:
\begin{widetext}
\begin{equation}
m_{\chi}~ \frac{[1-m_b^2/m_\chi^2]^{1/4}}{[1-(2m_{\chi})^2/m_A^2]}
\gsim 7.4 ~ {\rm GeV} \left(\frac{m_A}{90 \; {\rm GeV}} \right)^2 \left(\frac{35}{\tan \beta}\right)
\left(\frac{0.12}{a_1^2 a_3^2}\right)^{\frac{1}{2}}
\left(\frac{0.12}{(\Omega_{CDM} h^2)_{\rm max}}\right)^{\frac{1}{2}}.
\label{ma}
\end{equation}
\end{widetext}
Here we have taken as default value for
$(\Omega_{CDM} h^2)_{\rm max}$ the
numerical value which represents the 2$\sigma$ upper bound to $(\Omega_{CDM}
h^2)_{\rm max}$ derived from the results of Ref. \cite{wmap}. For
$\epsilon_b$ we have used a value which is representative of the typical range
obtained numerically in our model:
$\epsilon_b = -0.08$.

Eq. (\ref{ma}), already derived in Ref. \cite{lowneu},
is written here in a form which
shows more explicitly how  the lower limit on  $m_{\chi}$ depends on the various model
parameters. Notice in particular that this lower bound scales (roughly) as $m_A^2$ and
$(\tan \beta)^{-1}$.
It is obvious that the precise value for the lower limit
has however to be ascertained by numerical evaluations which take into
account all the intricate interferences of the various physical constraints
over the model parameters.

The right--hand--side of Eq. (\ref{ma}) can be expressed completely in terms of the independent parameters
of our SUSY model. In fact, by using Eq. (\ref{approx2}) at large $\tan \beta$ ($\sin \beta \simeq 1$)
and taking into account that, because of Eq. (\ref{hierarchy1}),  $a_1^2 \simeq 1 - a_3^2$, we can rewrite $a_1^2 a_3^2$ as:
\begin{equation}
a_1^2 a_3^2 \simeq \frac{\sin^2 \theta_W \; m^2_Z \; \mu^2}{(\mu^2 + \sin^2 \theta_W \; m^2_Z)^2} \simeq  \frac{0.19 \; \mu^2_{100}}{(\mu^2_{100} + 0.19)^2}.
\label{a1a3}
\end{equation}
From this formula and the LEP lower bound $|\mu| \gsim$ 100 GeV, we obtain
$(a_1^2 a_3^2)_{\rm max} \lsim 0.13$.

An upper limit on $a_3^2$ and then on the product $a_1^2 a_3^2$ is also placed
by the upper bound on the width for the $Z$--boson decay into a light neutralino
pair. This decay width is given by \cite{giapponese,barbieri}:
\begin{equation}
\Gamma(Z \rightarrow \chi \chi) = \frac{1}{12\pi}\frac{G_F}{\sqrt{2}} M^3_Z  [1-(2m_{\chi})^2/M_Z^2]^{3/2}
(a^2_3 - a^2_4)^2.
\label{barb}
\end{equation}
Taking into account that $a^2_3 \gg a^2_4$:
\begin{equation}
\Gamma(Z \rightarrow \chi \chi) = 166  \; {\rm MeV} \; [1-(2m_{\chi})^2/M_Z^2]^{3/2} \; a^4_3.
\label{barb2}
\end{equation}
Denoting by $\Gamma(Z \rightarrow \chi \chi)_{\rm ub}$ the upper bound to the invisible fraction of the $Z$--decay width,
we finally obtain:
\begin{equation}
a^2_3 \lsim
\left(\frac{\Gamma(Z \rightarrow \chi \chi)_{\rm ub}}{154 \; {\rm MeV}} \right)^{1/2},
\label{upper}
\end{equation}

\noindent
where we have used for the neutralino mass the value $m_{\chi} \simeq$ 10 GeV.

If we take conservatively  $\Gamma(Z \rightarrow \chi \chi) <$ 3 MeV
\cite{aleph05,pdg}, from Eq. (\ref{upper}) we find  $a_1^2 a_3^2 \lsim$ 0.12, a value
which is extremely close to the upper bound   $(a_1^2 a_3^2)_{\rm max}$= 0.13  derived above from Eq. (\ref{a1a3})
and the experimental lower limit on $|\mu|$.  The value $a_1^2 a_3^2 =$ 0.12 is the reference value
for   $a_1^2 a_3^2$ employed in Eq. (\ref{ma}).
Notice that the upper bound on $a_3^2$, placed
by the invisible width for $Z \rightarrow \chi + \chi$, scales
with the square root of the upper limit on this quantity (see Eq. (\ref{upper}));
furthermore, the lower bound
on $m_{\chi}$ scales as $(a_1^2 a_3^2)^{-1/2}$ (see Eq. (\ref{ma})).
Thus, the lower limit on the neutralino mass is only very mildly dependent on the actual value
of the upper bound on  $\Gamma(Z \rightarrow \chi \chi)$. For instance, taking
 $\Gamma(Z \rightarrow \chi \chi) <$ 2 MeV instead of $\Gamma(Z \rightarrow \chi \chi) <$ 3 MeV, would increase the lower bound on  $m_{\chi}$ by a mere 10\% \cite{dreiner}.

The properties of  the very light neutralinos of cosmological interest
considered in this Section delineate a specific scenario hereby denoted as
{\bf Scenario $\mathcal{A}$} \cite{lhc1}.
 Its main features are strongly determined
  by the requirement that the neutralino relic abundance satisfies the cosmological
  bound $\Omega_{\chi} h^2 \leq (\Omega_{CDM} h^2)_{\rm max}$. From the approximate formula
  in Eq. (\ref{eq:omega0}) one finds that: i) $m_A$ must be light,
  90 GeV $\leq m_A \lsim (200-300)$ GeV
 (90 GeV being the lower bound from LEP searches); 
  ii)  $\tan \beta$ has to be large: $\tan \beta$ = 20--45,
  iii) the  $\tilde {B} - \tilde H_1^{\circ}$ mixing needs to be sizeable, which in
  turn implies small values of $\mu$: $|\mu| \sim (100-200)$ GeV (see Eq. (\ref{a1a3})).
 As will be discussed in Sect. \ref{bsdecay},  the trilinear coupling is
constrained, for neutralinos lighter than 10 GeV, to be in the interval $|A| \lsim 0.6$, because of the
upper bound to $BR(B_s \rightarrow {\mu}^{+} + {\mu}^{-})$. 
Eq. (\ref{ma}) shows that in
Scenario {$\mathcal{A}$} we expect a lower bound on the neutralino mass of
the order of 7.5 GeV, if the parameter space which defines this Scenario
is allowed by the bounds on Higgs searches and B--physics. We
will show in the next Sections that this result actually holds.

\subsection{ Scenario {$\mathcal{B}$}}
\label{sec:scenarioB}

When $m_A \gsim (200-300)$ GeV, the cosmological lower bound on
$\sigmavint$ can be satisfied by a pair annihilation process
which proceeds through an efficient stau--exchange contribution (in the
{\it t, u} channels). This requires that: (i) the stau mass
$m_{\tilde{\tau}}$ is sufficiently light, $m_{\tilde{\tau}} \sim$ 90
GeV (notice that the current experimental limit is $m_{\tilde{\tau}}
\sim$ 87 GeV) and (ii) $\chi$ is a very pure Bino ({\it i.e.} $(1 -
a^2_1) \sim {\cal O}(10^{-2})$). 

The requirement (i) sets a constraint on the quantity
$|\mu| \tan \beta$, because the
experimental lower bounds on the sneutrino mass and on the charged
slepton masses of the first two families imply a lower bound on the
soft slepton mass: $m_{\tilde{l}} \gsim$ 115 GeV. Thus, in order to make the
request $m_{\tilde{\tau}} \sim$ 90 GeV compatible with $m_{\tilde{l}}
\gsim$ 115 GeV, it is necessary that the off-diagonal terms of the
sleptonic mass matrix in the eigenstate basis, which are proportional
to $\mu \tan \beta$, are large. Numerically, one finds $|\mu| \tan
\beta \sim$ 5000 GeV.
On the other side, the condition (ii) requires that $|a_3/a_1| \lsim 10^{-1}$, {\it i.e.},
according to  Eq. (\ref{approx2}),
$|a_3/a_1| \simeq \sin\theta_W \sin\beta \, (M_Z/\mu) \lsim 10^{-1}$. Combining this last expression with the condition $|\mu| \tan \beta \sim$ 5000 GeV, one finds that $|\mu|$ and $\tan \beta$ are bounded
by: $|\mu| \gsim$ 500 GeV, $\tan \beta \lsim$ 10. These bounds are somewhat weaker for
 values of the neutralino mass larger than $\sim$ 15--18 GeV.

The previous arguments lead us to introduce 
{\bf Scenario $\mathcal{B}$} \cite{lhc1}, identified by the following sector of the supersymmetric
parameter space: $M_1 \sim$ 25 GeV, $|\mu| \gsim$ 500 GeV, $\tan \beta
\lsim$ 10; $m_{\tilde{l}} \gsim (100 - 200)$ GeV, $-2.5 \lsim A \lsim +2.5$;
the other supersymmetric parameters are not {\it a priori} fixed.
 Within this scenario it
follows from Eqs. (\ref{approx1}--\ref{approx3}) that the following hierarchy holds
for the coefficients $a_i$:
\begin{equation}
|a_1| \gg |a_2|, |a_3|, |a_4|\, .
\label{hierarchy2}
\end{equation}
As derived in Ref. \cite{lowneu} the cosmological bound
$\Omega_{\chi} h^2 \leq (\Omega_{CDM} h^2)_{\rm max}$ provides
the lower bound  $m_{\chi} \gsim 22$ GeV,
whose scaling law in terms of  the stau mass
and $(\Omega_{CDM} h^2)_{\rm max}$ is approximately given by:
\begin{widetext}
\begin{equation}
m_{\chi} [1-m_{\tau}^2/m_\chi^2]^{1/4} \gsim 
22 \; {\rm GeV} \;
\left( \frac{m_{\tilde{\tau}}}{90 \; {\rm GeV}} \right)^2 \left(\frac{0.12}{(\Omega_{CDM} h^2)_{\rm max}}\right).
\label{tau}
\end{equation}
\end{widetext}

In general, one has conservatively to retain as a lower bound to
$m_{\chi}$ the smaller of the two lower limits given separately in
Eq. (\ref{ma}) and in Eq. (\ref{tau}). From these equations one finds
that the lower bound of Eq. (\ref{ma}) is less stringent than the one
of Eq. (\ref{tau}) as long as $m_A \lsim 2 m_{\tilde{\tau}}$. Due to
the present experimental bounds on $m_A, \tan \beta$ and $m_{\tilde{\tau}}$ the
lower absolute bound is the one derived from Eq. (\ref{ma}).
We parenthetically note that the
lower limits $m_{\chi} \gsim (15-18)$ GeV found in
Refs. \cite{hooper,boudjema} are due to the assumption that
$m_A$ is very large ($m_A \sim 1$ TeV).

\section{Numerical analysis of the LNM parameter space}
\label{scanning}

In the present paper we are interested in discussing neutralinos with
very light masses. We will therefore concentrate on Scenario $\cal A$
only, and in our numerical analyses we perform a scanning of the supersymmetric 
parameter space dedicated to this scenario. We will denote this sector of the
LNM parameter space as LNM--$\cal A$.

The ranges of the MSSM parameters, appropriately narrowed in order to explore this scenario are:
$10 \leq \tan \beta \leq 50$,
$100 \, {\rm GeV} \leq \mu \leq 150 \, {\rm GeV}$,
$5 \, {\rm GeV} \leq M_1 \leq 50 \, {\rm GeV}$,
$100 \, {\rm GeV} \leq M_2 \leq 1000 \, {\rm GeV}$,
$250 \, {\rm GeV} \leq m_{\tilde q} \leq 1000 \, {\rm GeV }$,
$100 \, {\rm GeV} \leq m_{\tilde l} \leq 3000 \, {\rm GeV }$, 
$90\, {\rm GeV }\leq m_A \leq 120 \, {\rm GeV }$,
$0 \leq A \leq 1$.

In our scenario the trilinear coupling $A$ and the
$\mu$ parameter are both always positive. This is actually
due to the interplay between the constraints on
$BR(b\rightarrow s \gamma)$ (which requires $\mu A>0$) and on
$a_{\mu}$ (which requires $\mu>0$). In all the plots shown in
the paper, all the experimental bounds discussed in Sect.
\ref{sec:model} are applied: {\em i.e.} invisible $Z$--width,
Higgs searches at LEP and Tevatron, $BR(b \rightarrow s \gamma)$,
muon anomalous magnetic moment $a_\mu$, $BR(B_s \rightarrow {\mu}^{+}{\mu}^{-})$,
$\Delta M_{B,s}$, $BR(t\rightarrow b H^+)$, and the upper bound on the
cosmological abundance $\Omega_\chi h^2$.
Whenever we refer to ``LNM--$\cal A$ scan", we intend the scan of the
parameters space defined above, implemented by the experimental bounds quoted
here.

As for the scan of the parameter space, we randomly sample the above intervals using a logarithmic scale. 
It is worth noticing here that, due to the non--trivial interplay of the
different constraints on physical masses and couplings, the typical
success rate of our sampling
for obtaining neutralinos with mass less than 10 GeV
 is of order $10^{-5}$--$10^{-6}$. In
particular, in order to populate the scatter plots with a sizable
number of points and in a uniform way, we have subdivided the above
ranges in smaller intervals for the neutralino mass $m_{\chi}\simeq
M_1$ and run our code until a similar number of allowed points were
found in each sub-range. Less focused
scans of the parameter space may fail to find allowed configurations,
especially in the lower range of $m_{\chi}$,
as seems to be the case with some analyses in the literature \cite{belanger}.

All the numerical results shown in the paper refer to this special
LNM--$\cal A$ scan. The only exception will be 
Fig. \ref{fig:scat_mchi_sigmac_full}, where a more general scan of the
effMSSM will be presented. In that (unique) case, the parameters
will be varied in the following intervals:
$1 \leq \tan \beta \leq 50$,
$100 \, {\rm GeV} \leq |\mu| \leq 1000 \, {\rm GeV}$,
$5 \, {\rm GeV} \leq M_1 \leq \min(100,0.5 M_2) \, {\rm GeV}$,
$100 \, {\rm GeV} \leq M_2 \leq 1000 \, {\rm GeV}$,
$80 \, {\rm GeV} \leq m_{\tilde q} \leq 3000 \, {\rm GeV }$,
$80 \, {\rm GeV} \leq m_{\tilde l} \leq 3000 \, {\rm GeV }$, 
$90\, {\rm GeV }\leq m_A \leq 1000 \, {\rm GeV }$,
$-1 \leq A \leq 1$.
The scan of Fig. \ref{fig:scat_mchi_sigmac_full} will therefore
include both Scenario $\cal A$ and Scenario $\cal B$, as well as more
general scenarios, with heavier neutralinos. Within the scan at higher
neutralino masses described above we have imposed the condition
$m_{NLSP}>1.05\, m_{\chi}$ (with NLSP=sfermions,charginos) in order to
remove configurations where the relic abundance is determined by
coannihilations between the neutralino and the Next--to--Lightest SUSY
Particle (NLSP). Notice that in an effective MSSM coannihilations are
due to accidental degeneracies between uncorrelated parameters. This
is at variance with the SUGRA scenario, where strong correlations
among the mass of the neutralino and of other SUSY particles are
expected in particular regions of the parameter space.

Now it is convenient to have a first look at a scatter plot for the neutralino population
within the LNM--$\cal A$. This is provided  by Fig. \ref{fig:scat_mchi_tanb} where the scatter plot is represented in the plane $m_{\chi}-\tan \beta$.
In evaluating this scatter plot all constraints specifically mentioned
in Sect. \ref{sec:model} (including the upper bound on $BR(B_s \rightarrow {\mu}^{+} {\mu}^{-})$
and the cosmological upper bound $\Omega_{\chi} h^2 \leq (\Omega_{CDM} h^2)_{\rm max} = 0.12$) have been applied.
Notice that  $\Omega_{\chi} h^2$ has been evaluated using its full expression, and not simply
with its approximate version given in Eq. (\ref{eq:omega0}). This figure shows how accurate is the bound given in Eq. (\ref{ma}), which is represented by
the (red) solid line. From this figure  it turns out that
 the lower bound on the neutralino mass is
 $m_{\chi} \gsim$ 7.5 GeV, this value being obtained when $\tan \beta \simeq$ 40
 and $m_A \simeq$ 90 GeV.
We stress that the updated constraint on $BR(B_s \rightarrow \mu^+ \mu^-)$ induces
 only a very
slight modification in the neutralino mass lower bound of 6--7 GeV determined in
Ref. \cite{lowneu}. This constraint will be further discussed in Sect. \ref{bsdecay}.

In Fig. \ref{fig:scat_mchi_gammaz} we give a scatter plot for
$\Gamma(Z \rightarrow \chi \chi)$ versus $m_{\chi}$. From this plot
one sees that actually the lower bound on the neutralino mass changes very little
when the upper bound on the invisible fraction of the Z-boson width, shown by the horizontal
solid line, is
decreased from 3 MeV to about 2 MeV, as previously argued. Somewhat below this value
the impact of the invisible Z--width on the lower bound of variation of
$m_{\chi}$  may be substantial.

\section{Neutralino--nucleon cross--section}
\label{sec:cross}

In the present paper we analyze the results of present experiments
searching for direct detection of DM particles,  under the hypothesis that WIMPs have
a dominant coherent interaction with the detector nuclei. This is the case for neutralinos.
Once a specific distribution is assumed for the WIMPs in the halo, the WIMP--nucleus cross--section
can be immediately rewritten in terms of the WIMP--nucleon cross--section; this is then  the central
quantity to be analyzed.

Thus, we turn now to an approximate evaluation of the neutralino--nucleon cross--section
in the case where the interaction process is due to exchange of the lighter CP--even
neutral Higgs boson $h$.
From the formulae in Refs.\cite{bdfs2,uncert2} one obtains:
\begin{equation}
\sigma_{\rm scalar}^{(\rm nucleon)} \simeq \frac {8 G_F^2} {\pi} M_Z^2
m_{\rm red}^2 \;
\frac{F_h^2 I_h^2}{m_h^4}\, ,
\label{eq:sigmasc}
\end{equation}
where:
\begin{eqnarray}
F_h &=& (-a_1 \sin \theta_W+a_2 \cos \theta_W) (a_3 \sin \alpha + a_4 \cos \alpha)\, 
\nonumber\\
I_{h}&=&\sum_q k_q^h m_q \langle N|\bar{q} q |N\rangle \, .
\label{eq:I}
\end{eqnarray}
The matrix elements $\langle N|\bar{q}q|N \rangle$ are meant over the nucleonic
state, the angle $\alpha$ rotates $H_1^{(0)}$ and $H_2^{(0)}$ into $h$ and
$H$, and the coefficients $k_q^h$ are given by:
\begin{eqnarray}
k_{u{\rm -type}}^h &=& ~\cos\alpha / \sin\beta  \, , \nonumber\\
k_{d{\rm -type}}^h &=& - \sin\alpha / \cos\beta -\epsilon_d \cos(\alpha-\beta) \tan\beta \, ,
\label{eq:k}
\end{eqnarray}
for the up--type and down--type quarks, respectively; $\epsilon_d$ has already been introduced in Eq. (\ref{eq:omega0}) for the case of the $b$ quark.

Keeping the dominant terms (couplings of the Higgs boson $h$ with the $d$--type quarks, $\alpha \simeq \pi$/2), one has:
\begin{eqnarray}
I_{h}&\simeq& k_{d{\rm -type}}^h [m_d \langle N|\bar{d} d |N\rangle + m_s \langle N|\bar{s} s |N\rangle +
m_b \langle N|\bar{b} b |N\rangle] \nonumber \\
&\simeq&  - \tan \beta \; g_d \, ,
\label{eq:Iappr}
\end{eqnarray}
where:
\begin{equation}
g_d \equiv [m_d \langle N|\bar{d} d |N\rangle + m_s \langle N|\bar{s} s |N\rangle +
m_b \langle N|\bar{b} b |N\rangle].
\label{eq:gd}
\end{equation}
Thus:
\begin{equation}
\sigma_{\rm scalar}^{(\rm nucleon)} \simeq 6.8  \times 10^{-7} a_1^2 a_3^2 \tan^2 \beta \; \frac{g_d^2}{m_h^4} ,
\label{sel1}
\end{equation}
or:
\begin{widetext}
\begin{equation}
\sigma_{\rm scalar}^{(\rm nucleon)} \simeq 5.3  \times 10^{-41} \; {\rm cm^2} \;
\left(\frac{a_1^2 a_3^2}{0.13} \right)
\left(\frac{\tan \beta}{35} \right)^2
\left(\frac{90 \; {\rm GeV}}{m_h} \right)^4
\left(\frac{g_d}{290 ~ {\rm MeV}} \right)^2.
\label{sel2}
\end{equation}
\end{widetext}

In this expression we have used as {\it reference} value for $g_d$ the value 
$g_{d,\rm ref} = 290$ MeV employed in our previous
papers \cite{zoom,inter}. We recall that this quantity is affected by large uncertainties \cite{uncert2} with
$\left({g_{d,\rm max}}/{g_{d,\rm ref}}\right)^2 = 3.0$ and 
$\left({g_{d,\rm min}}/{g_{d,\rm ref}}\right)^2 = 0.12$ \cite{inter}.
Our reference value $g_{d,\rm ref} = 290$ MeV is larger by  a factor 1.5 than the central value of Ref. \cite{efo}, frequently used in literature (see for instance Ref. \cite{kuflik}).

By employing Eq. (\ref{eq:omega0}) and Eq. (\ref{sel1}) we obtain:

\begin{widetext}
\begin{equation}
(\Omega_{\chi} h^2) \; \sigma_{\rm scalar}^{(\rm nucleon)} \simeq 3.3 \times 10^{-39} \;{\rm cm^2} \;
g_d^2 \;
\frac{[1-(2m_{\chi})^2/m_A^2]^2}{m_{\chi}^2~[1-m_b^2/m_\chi^2]^{1/2}} \; \frac{1}{(1 + \epsilon_b)^2}.
\label{prod}
\end{equation}
\end{widetext}

From this expression we find that any neutralino configuration, {\it whose relic abundance stays in the
cosmological range for CDM} ({
\it i.e.} $(\Omega_{CDM} h^2)_{\rm min} \leq \Omega_{\chi} h^2 \leq (\Omega_{CDM} h^2)_{\rm max}$ with
$(\Omega_{CDM} h^2)_{\rm min} = 0.098$ and  $(\Omega_{CDM} h^2)_{\rm max} = 0.12$) and passes all particle--physics constraints,  has an elastic neutralino--nucleon cross--section of order:
\begin{widetext}
\begin{equation}
\sigma_{\rm scalar}^{(\rm nucleon)} \simeq (2.7 - 3.4) \times 10^{-41} \; {\rm cm^2}  \;
\left(\frac{g_d}{290 ~ {\rm MeV}} \right)^2
\frac{[1-(2m_{\chi})^2/m_A^2]^2}{(m_{\chi}/(10 \; {\rm GeV})^2 \; [1- m_b^2/m_\chi^2]^{1/2}}.
\label{bound}
\end{equation}
\end{widetext}

A few comments are in order here:

i)  The elastic cross--section $\sigma_{\rm scalar}^{(\rm nucleon)}$ is affected by large uncertainties
because of the uncertainties inherent in the effective Higgs--quark coupling constant $g_d$ \cite{uncert2}.
Actually,  $\sigma_{\rm scalar}^{(\rm nucleon)}$ is subject to an increase
by a factor of 3.0 or to a decrease by a factor of 8.6 \cite{inter}, as 
commented above;

ii) Eq. (\ref{bound}) shows that $\sigma_{\rm scalar}^{(\rm nucleon)}$  scales roughly as $(m_{\chi})^{-2}$ for the
 range of neutralino masses considered here;

iii) To establish the range of $m_{\chi}$ to which Eq. (\ref{bound}) applies, one simply has to evaluate
the lower bound $m_{\chi}$ by using Eq. (\ref{ma});

iv) The bounds, set by particle--physics measurements,
on the two  parameters $\tan \beta$ and $m_A$ have a strong impact on the lower bound of
$m_{\chi}$ (see Eq. (\ref{ma})), but are either uninfluential (in the case of $\tan \beta$) or only marginally influent (in the case
of $m_A$) in the estimate of  $\sigma_{\rm scalar}^{(\rm nucleon)}$.

Furthermore, we wish to notice that also the situation when relic
neutralinos only provide a fraction of the CDM abundance is of great
importance. Indeed, as shown in Ref. \cite{rescaling}, in a direct
detection experiment, relic neutralinos, whose relic abundance does
not saturate the CDM abundance (that is, with $\Omega_{\chi} h^2 \leq
(\Omega_{CDM} h^2)_{\rm min}$), have a response larger than
neutralinos of higher relic abundance. This property is due to the
fact that, for subdominant neutralinos, the direct detection rate has
to include a factor which appropriately depletes the value of the
local DM density $\rho_0$ when $\Omega_{\chi} h^2 \leq (\Omega_{CDM}
h^2)_{\rm min}$. This {\it rescaling} factor $\xi = \rho_{\chi} /
\rho_0$ is conveniently taken as $\xi = {\rm min}\{1, \Omega_{\chi}
h^2/(\Omega_{CDM} h^2)_{\rm min}\}$ \cite{gaisser}. Thus, effectively
the relevant quantity to be inserted in the detection rate is not
simply $\sigma_{\rm scalar}^{(\rm nucleon)}$ but rather $\xi
\sigma_{\rm scalar}^{(\rm nucleon)}$.  If one performs a scanning of
the supersymmetric parameter space, it turns out that, at fixed
$m_{\chi}$, the quantity $\xi \sigma_{\rm scalar}^{(\rm nucleon)}$,
when plotted including configurations with a subdominant relic
density, can provide larger values than $\sigma_{\rm scalar}^{(\rm
  nucleon)}$ when the latter is plotted in the case when only
neutralinos providing the observed DM density are included.  This
feature is manifest in the numerical results presented hereafter.
These important rescaling properties are often overlooked in current
phenomenological analyses of experimental data.

\section{Constraints on supersymmetric parameters from the Fermilab Tevatron collider and the B-factories}
\label{sec:constraints}

Now we discuss some relevant particle--physics measurements for which there have been sizable
improvements recently, or
which might become important in the near future as new data become available.
We consider how each of the present experimental results on searches for new physics
at the Tevatron and at the B--factories can impact on the LNM--$\cal A$ 
by putting constraints mainly on the two crucial parameters $m_A$ and $\tan \beta$.
Once these bounds are established, we determine how these limits reflect on the lower bound
for the neutralino mass and consequently on the neutralino--nucleon cross--section. Our analysis
is performed analytically and numerically individually for each measurement, since the
results of the various particle--physics experiments  do not share the same level of
reliability; actually,  some of them are still presented by the experimental collaborations under
 the form of preliminary reports. Thus, in some case it is still premature to enforce the
corresponding constraints at the present stage, though these might possibly become relevant
in the future.

\subsection{Search for the rare decay $B_s \rightarrow \mu^+ + \mu^-$  at the Tevatron}
\label{bsdecay}

The SUSY contributions to the branching ratio for the decay $B_s \rightarrow \mu^+ + \mu^-$
are  very sensitive to  $\tan \beta$, since for high values of
this parameter they behave as  $\tan^6 \beta$  \cite{bobeth,arnowitt,buras,eos}. Thus, the experimental upper bound on the branching ratio for $B_s \rightarrow \mu^+ + \mu^-$ can potentially put strict constraints on the elastic neutralino--nucleon cross--section when this proceeds through a Higgs--exchange, as is the case for the cross--section in Eq. (\ref{sel2}). However, we wish to stress here that the actual impact of these constraints depends dramatically on the specific SUSY model.

To clarify  this point, we start discussing  the features of the supersymmetric contributions to
the branching ratio for $B_s \rightarrow \mu^+ + \mu^-$ which actually go like $\tan^6 \beta$,
as derivable for instance from
Eqs.(1)--(2)  of Ref. \cite{arnowitt} (notice however that in the numerical evaluations of
$BR(B_s \rightarrow \mu^+ \mu^-)$ reported later on, all supersymmetric contributions are
included as given in Ref. \cite{bobeth}). The dominant contribution which behaves as $\tan^6 \beta$ writes:
\begin{widetext}
\begin{eqnarray}
BR^{(6)}(B_s \rightarrow \mu^+ \mu^-) &\simeq& \frac{1}{2^{12} \pi^3} \;
\frac{G^2_F \alpha^2}{\sin^4 \theta_W} \; \tau_B M^5_B f^2_{B_s}
\left(\frac{m_{\mu} m_t m_{\chi_{\pm}}}{m^2_W m^2_A}\right)^2
\left[1 + \left(\frac{m_b - m_s}{m_b + m_s}\right)^2 \right]
\sin^2(2 \theta_{\tilde t}) \times \nonumber \\
& & \times  |V_{tb}|^2 \; |V_{ts}|^2 \; \tan^6 \beta \;
\left[D(m^2_{\tilde{t}_2}/{\mu^2}) - D(m^2_{\tilde{t}_1}/{\mu^2})\right]^2,
\label{mumu}
\end{eqnarray}
\end{widetext}
where $\tau_B$ is the $B$ meson mean life, $M_B$ is its mass, $f_{B_s}$ is the $B_s$ decay constant,
$m_t, m_b, m_s$ are the masses of the top, bottom and strange quarks, respectively; $V_{tb}$ and
$V_{ts}$ are elements of the Cabibbo--Kobayashi--Maskawa matrix and the function $D(x)$ is defined
as  $D(x) = x \; \log(x)/(1 - x)$.
The structure of $BR^{(6)}(B_s \rightarrow \mu^+ \mu^-)$ in
Eq. (\ref{mumu}) is due to the fact that the relevant amplitudes contain a one--loop insertion
on a quark line, the loop being formed by a chargino (of mass $m_{\chi^{\pm}} \sim |\mu|$ in
our models) and a stop
whose mass eigenvalues are denoted as $m^2_{\tilde{t}_i}$ (i = 1, 2); $\theta_{\tilde t}$ is the rotation angle which comes out
when the stop squared-mass matrix is diagonalized.
If in Eq. (\ref{mumu}) we insert  the values:
$\tau_B = (1.47 \pm 0.02) \times 10^{-12}$ sec, $M_B$ = 5.37 GeV, $f_{B_s} = (210 \pm 30)$ MeV, $|V_{tb}| = 0.88 \pm 0.07$
and $|V_{ts}| = (38.7 \pm 2.1) \times 10^{-3}$, we obtain (by using the central values for the various quantities):

\begin{widetext}
\begin{eqnarray}
BR^{(6)}(B_s \rightarrow \mu^+ + \mu^-) &\simeq& 2.70 \times 10^{-6} \; \sin^2(2 \theta_{\tilde t}) \;
\left(\frac{m_{\chi^{\pm}}}{110 \; {\rm GeV}}\right)^2
\left(\frac{90 \; {\rm GeV}}{m_A}\right)^4
\left(\frac{\tan \beta}{35}\right)^6 \times \nonumber \\
& & \times
\left[D(m^2_{\tilde{t}_2}/{\mu^2}) - D(m^2_{\tilde{t}_1}/{\mu^2})\right]^2.
\label{mumu2}
\end{eqnarray}
\end{widetext}

\noindent
The uncertainty in the numerical factor in front of the RHS of this equation is of about 40\%.

Since the purpose of the present discussion is essentially illustrative to show in which features the size of
$BR^{(6)}$ in our LNM--$\cal A$ may differ from its size in SUGRA models, we proceed to some approximations.
In the scenario {$\mathcal{A}$} of our model (see Sect. \ref{sec:model})
$|\mu|$ is typically small
(close to the current LEP lower bound $|\mu| \gsim$ 100 GeV) at variance with what occurs in SUGRA models,
where the SUSY--breaking implies larger values of $|\mu|$; then here
 the ratios $m^2_{\tilde{t}_{1,2}}/{\mu^2}$ are large:  $m^2_{\tilde{t}_{1,2}}/{\mu^2} \gg 1$.

Since $D(x) \simeq - \log(x)$ when $x \gg 1$, one obtains:
\begin{eqnarray}
\left[D(m^2_{\tilde{t}_2}/{\mu^2}) - D(m^2_{\tilde{t}_1}/{\mu^2})\right]^2 \simeq 
\left[ \log ({m^2_{\tilde{t}_2}}/{m^2_{\tilde{t}_1}}) \right]^2 \, .
\label{mumu3}
\end{eqnarray}
Neglecting the contributions of the D-terms, the stop--mass eigenvalues  are approximately given by:
\begin{equation}
m^2_{\tilde{t}_{2,1}} = m^2_{\tilde q} + m^2_t \pm m_t (A \; m_{\tilde q} + \mu/\tan \beta).
\label{stopmass}
\end{equation}
In the scenario {$\mathcal{A}$}, where $\tan \beta$ is large and $|\mu|$ is small,
unless the trilinear coupling is practically null,
we have $|A| \gg |\mu|/(m_{\tilde q} \; \tan \beta$) and consequently:
\begin{equation}
m^2_{\tilde{t}_{2}}/m^2_{\tilde{t}_{1}} \simeq 1 + \frac{2 \; |A| \; m_t \; m_{\tilde q}}{m^2_{\tilde q} + m^2_t}.
\label{stopratio}
\end{equation}
By inserting Eq. (\ref{mumu3}) and Eq. (\ref{stopratio}) into  Eq. (\ref{mumu2}) and
taking into account that in our model $\sin^2(2 \theta_{\tilde t}) \simeq 1$,
we obtain:
\begin{widetext}
\begin{eqnarray}
BR^{(6)}(B_s \rightarrow \mu^+ \mu^-) &\simeq& 5.8 \times 10^{-8}  \;
\left(\frac{14 \; A \; m_t \; m_{\tilde q}}{m^2_{\tilde q} + m^2_t}\right)^2
\left(\frac{m_{\chi^{\pm}}}{110 \; {\rm GeV}}\right)^2
\left(\frac{90 \; {\rm GeV}}{m_A}\right)^4
\left(\frac{\tan \beta}{35}\right)^6,
\label{mumu4}
\end{eqnarray}
\end{widetext}
where the numerical coefficient in front of the right-hand-side is normalized
to the experimental upper bound at 95\% C.L.,
$BR(B_s \rightarrow \mu^+ \mu^-) \leq  5.8 \times 10^{-8}$ \cite{cdf_mumu}
(this is the latest published value by the CDF Collaboration;
a somewhat smaller value is reported in an unpublished CDF Public Note 9892 \cite{cdf_note}).

Thus, neutralino configurations with a trilinear coupling parameter:

\begin{eqnarray}
|A|  \lsim  \frac{1}{14} \; \frac{m^2_{\tilde q} + m^2_t}{m_t \; m_{\tilde q}} \lsim 0.36
\label{Apar}
\end{eqnarray}

\noindent
are compatible with the constraint imposed by the upper bound on the branching ratio of the
$B_s \rightarrow \mu^+ + \mu^-$ process.
In the last step we have taken $m_{\tilde q} \simeq$ 1 TeV.
We recall that, because of the uncertainties involved in the determination of the numerical factor in Eq. (\ref{mumu2}), the numerical coefficients in Eq. (\ref{Apar}) are affected by an uncertainty of about 20\%. 
In addition, the approximation of Eq. (\ref{mumu3}) is only partly valid, since in
our scenario the arguments $x_i=m^2_{\tilde{t}_{1,2}}/{\mu^2}$ of the $D(x)$ function are large but not exceedingly large and a further correction is at hand: a careful analysis shows that the value in Eq. (\ref{mumu4}) is actually reduced by a factor
$\eta = (0.75 \div 1)$, for $x_i = (10 \div 100)$. This implies that the range on
$|A|$ given in  Eq. (\ref{Apar}) can extend to values larger by a factor $\eta^{-1/2}$, {\em i.e.} $|A| \lsim (0.36 \div 0.42)$.

We will see now that the situation is even more favorable than the one depicted in 
Eq. (\ref{mumu4}), once the role of
the other SUSY contributions concurring to the
full calculation of  $BR(B_s \rightarrow \mu^+ \mu^-)$ is taken into account.
In Figs. \ref{fig:scat_mchi_bsmumu_contrib0} and \ref{fig:scat_tanb_bsmumu_contrib0} 
we show the absolute value of the Wilson coefficients for each SUSY contribution
to the $BR(B_s \rightarrow \mu^+ \mu^-)$ (colored points), compared to the full calculation 
of the dominant term (black points) (for expressions of these quantities see Refs. \cite{bobeth,arnowitt,buras,eos}), for our full scan in LNM--$\cal A$. The sign of each term is indicated in parenthesis as ``$[+]$'' and ``$[-]$''. We see that also some other terms (notably, the W--boson, the Higgs and the penguin diagrams) can contribute significantly, some with opposite signs, to the total branching ratio. The result of the full calculation when compared with the approximate expression of Eq. (\ref{mumu4}) is shown in Figs. \ref{fig:scat_mchi_bsmumu_contrib} and \ref{fig:scat_tanb_bsmumu_contrib}. For models
with light neutralinos, especially in the case of $m_\chi \lsim 10$ GeV, the full 
calculation can be smaller that the approximate one by up to a factor of three.

\begin{figure}[t]
\includegraphics[width=1.1\columnwidth,clip=true,bb=18 60 592 520]{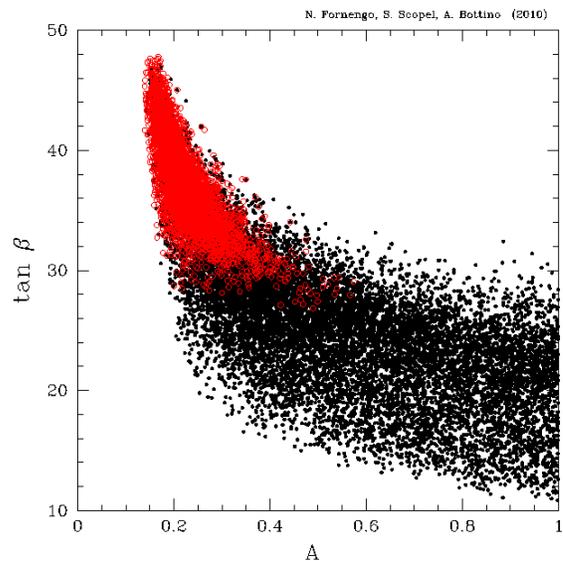}
\caption{Scatter plot which shows the correlation between 
$\tan\beta$ and the trilinear parameter $A$, for the
LNM--$\cal A$ scan. Black points refer to $m_\chi > 10$ GeV, red circles
to light neutralinos with $m_\chi \leq 10$ GeV}
\label{fig:scat_a_tanb}
\end{figure}

Fig. \ref{fig:scat_a_tanb} shows the correlation between $\tan\beta$
and the trilinear parameter $A$ in the LNM--$\cal A$ scan. We remind that here, as in all figures, the bound
$BR(B_s \rightarrow \mu^+ \mu^-) < 5.8 \times 10^{-8}$ is implemented. 
We see that, because of the competition among various contributions, the range of $A$ for light neutralinos (red circles) is wider than the one derived in Eq. (\ref{Apar}) and can extend up to about 0.6. 

Taking into account the requirements on $|A|$ used in deriving the previous
analytical approximations, {\em i.e.} $|A| \gg |\mu|/(m_{\tilde q} \; \tan \beta$), and
the upper bound of Eq. (\ref{Apar}), we find that $|A|$ has to satisfy the conditions:
\begin{eqnarray}
\frac{|\mu|}{m_{\tilde q} \; \tan \beta} \ll |A| \ll \frac{m_{\tilde q}}{m_t},
\label{rangeA}
\end{eqnarray}
{\it i.e.} a hierarchy which is naturally realized ({\it i.e.} no fine--tuning is involved) in our model where the values of the parameters are defined at the EW scale and not
induced by SUGRA conditions.

\begin{figure}[t]
\includegraphics[width=1.1\columnwidth,clip=true,bb=18 70 592 520]{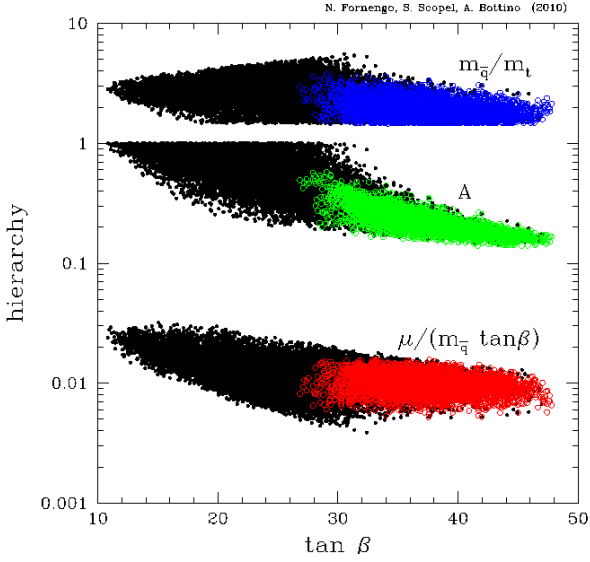}
\caption{Scatter plot which shows the hierarchy of Eq. (\ref{rangeA})
 in the LNM--$\cal A$ scan.
The three separate regions correspond, from top to bottom, to the numerical values for the three quantities $m_{\tilde q}/m_t$, $A$ and $\mu/(m_{\tilde q} \tan\beta)$.
 Coloured circles denote the neutralino configurations with $m_{\chi} \leq 10$ GeV.
}
\label{fig:scat_bsmumu_hierarchy}
\end{figure}

A demonstration of how this hierarchy is actually realized is provided by the numerical results in Fig. \ref{fig:scat_bsmumu_hierarchy}. The three separate regions correspond to the numerical values for the three quantities specified in the picture and in
Eq. (\ref{rangeA}). The (coloured) circles denote the neutralino configurations with $m_{\chi} \leq 10$ GeV.

\begin{figure}[t]
\includegraphics[width=1.1\columnwidth,clip=true,bb=18 60 592 520]{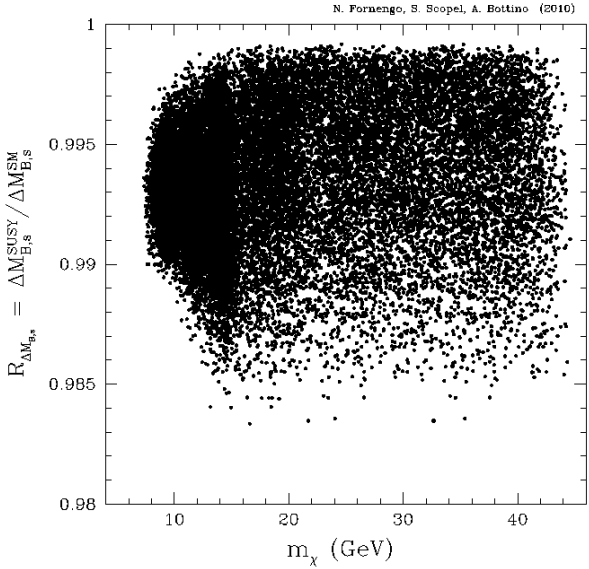}
\caption{Scatter plot of the ratio
$R_{\Delta M_{B,s}} = \Delta M^{SUSY}_{B,s}$/$\Delta M^{SM}_{B,s}$ as a function of
$m_{\chi}$, in the LNM--$\cal A$ scan.}
\label{fig:scat_mchi_deltamb}
\end{figure}

As discussed in the literature \cite{buras_delta,isidori} a supersymmetric contribution leading to an increase of  the decay rate of the process $B_s \rightarrow \mu^+ + \mu^-$ is correlated to a decrease of the difference
$\Delta M_{B,s} \equiv M_{B_s} - M_{\bar{B}_s}$, compared to the value expected in the Standard Model, $\Delta M^{SM}_{B,s}$.
In Fig. \ref{fig:scat_mchi_deltamb} we show  a scatter plot of the ratio  
$R_{\Delta M_{B,s}} = \Delta M^{SUSY}_{B,s}$/$\Delta M^{SM}_{B,s}$ as a function of
$m_{\chi}$ for the LNM--$\cal A$ scan. 
Taking into account that $R_{\Delta M_{B,s}} = 0.80 \pm 0.12$ \cite{isidori}, which
at 95\% C.L. implies an allowed range $0.57 < R_{\Delta M_{B,s}} < 1.03$,
one sees that the quantity $\Delta M_{B,s}$ does not
imply any additional constraint on the LNM parameter space.

To summarize the results of the previous discussion, we can say that in the LNM the upper
bound on the branching ratio of the $B_s \rightarrow \mu^+ + \mu^-$ process
 is not significantly constraining, due to the relatively small values of the chargino mass
and a small splitting between the two stop masses. These are situations which
are naturally obtained in our LNM, contrary to the situation that occurs in SUGRA--like models, as the one considered in Ref. \cite{nath}. It is then erroneous to interpret the results of Ref. \cite{nath} as valid also for a
generic effective MSSM with light neutralinos, as is sometimes done in recent literature
(see for instance Refs. \cite{belanger,gunion}).

\subsection{Search for neutral MSSM Higgs bosons at the Tevatron}
\label{searchneutralH}

%
\begin{figure}[t]
\includegraphics[width=1.1\columnwidth,clip=true,bb=18 190 592 650]{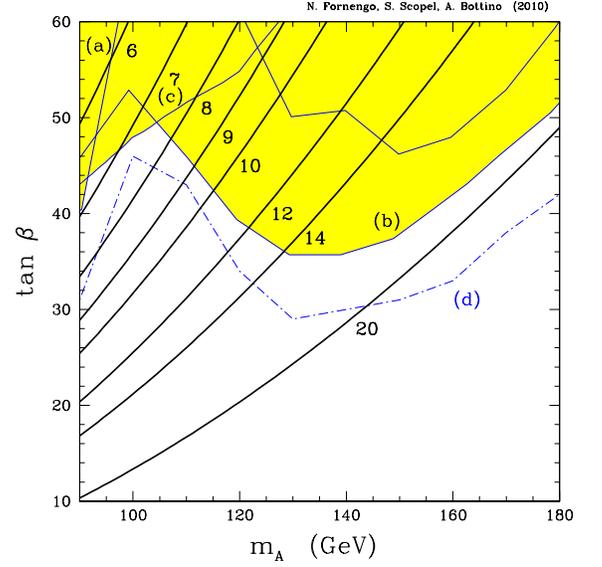}
\caption{Upper bounds in the $m_A$ -- $\tan\beta$ plane, derived from searches of the neutral Higgs boson at the Tevatron: line (a) is from Ref. \cite{abazov_tautau_2008},
line (b) from  Ref. \cite{aaltonen_tautau_2009}, line (c) from Ref. \cite{abazov_tautau_2009}.
The dot--dashed line (d) represents the preliminary bound given in Ref. \cite{tevnp}.
The solid bold lines labeled by numbers denote the cosmological bound 
$\Omega_\chi h^2 \leq (\Omega_{CDM} h^2)_{\rm max}$
for a neutralino whose mass
is given by the corresponding number (in units of GeV), as obtained
by Eqs. (\ref{eq:omega0}, \ref{ma}) with $\epsilon_b = -0.08$ and 
$(\Omega_{CDM} h^2)_{\rm max}=0.12$. For any given neutralino mass, the
allowed region is above the corresponding line.}
\label{fig:omega_tanb_higgs}
\end{figure}
\begin{figure}[t]
\includegraphics[width=1.1\columnwidth,clip=true,bb=18 60 592 520]{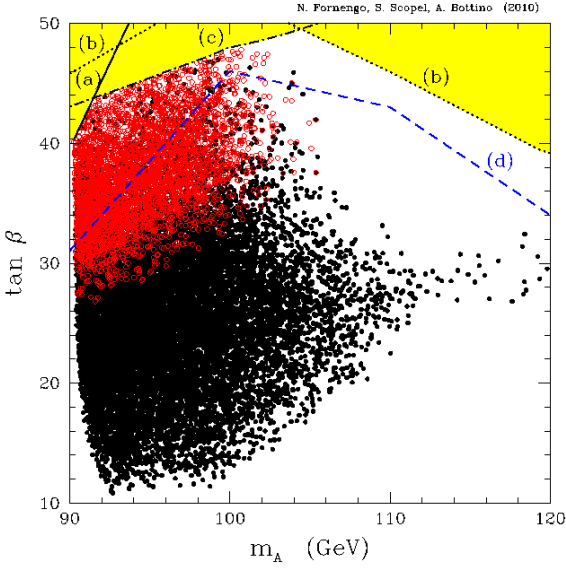}
\caption{Upper bounds in the $m_A$ -- $\tan\beta$ plane, derived from searches of the neutral Higgs boson at the Tevatron: line (a) is from Ref. \cite{abazov_tautau_2008},
line (b) from  Ref. \cite{aaltonen_tautau_2009}, line (c) from Ref. \cite{abazov_tautau_2009}.
The dot--dashed line (d) represents the preliminary bound given in Ref. \cite{tevnp}.
The scatter plot refers to the light neutralino population of the LNM--$\cal A$ scan.
Black points stand for $m_\chi > 10$ GeV, while the red circles for
$m_\chi \leq 10$ GeV.
}
\label{fig:scat_higgs_tanb}
\end{figure}
\begin{figure}[t]
\includegraphics[width=1.1\columnwidth,clip=true,bb=18 190 592 650]{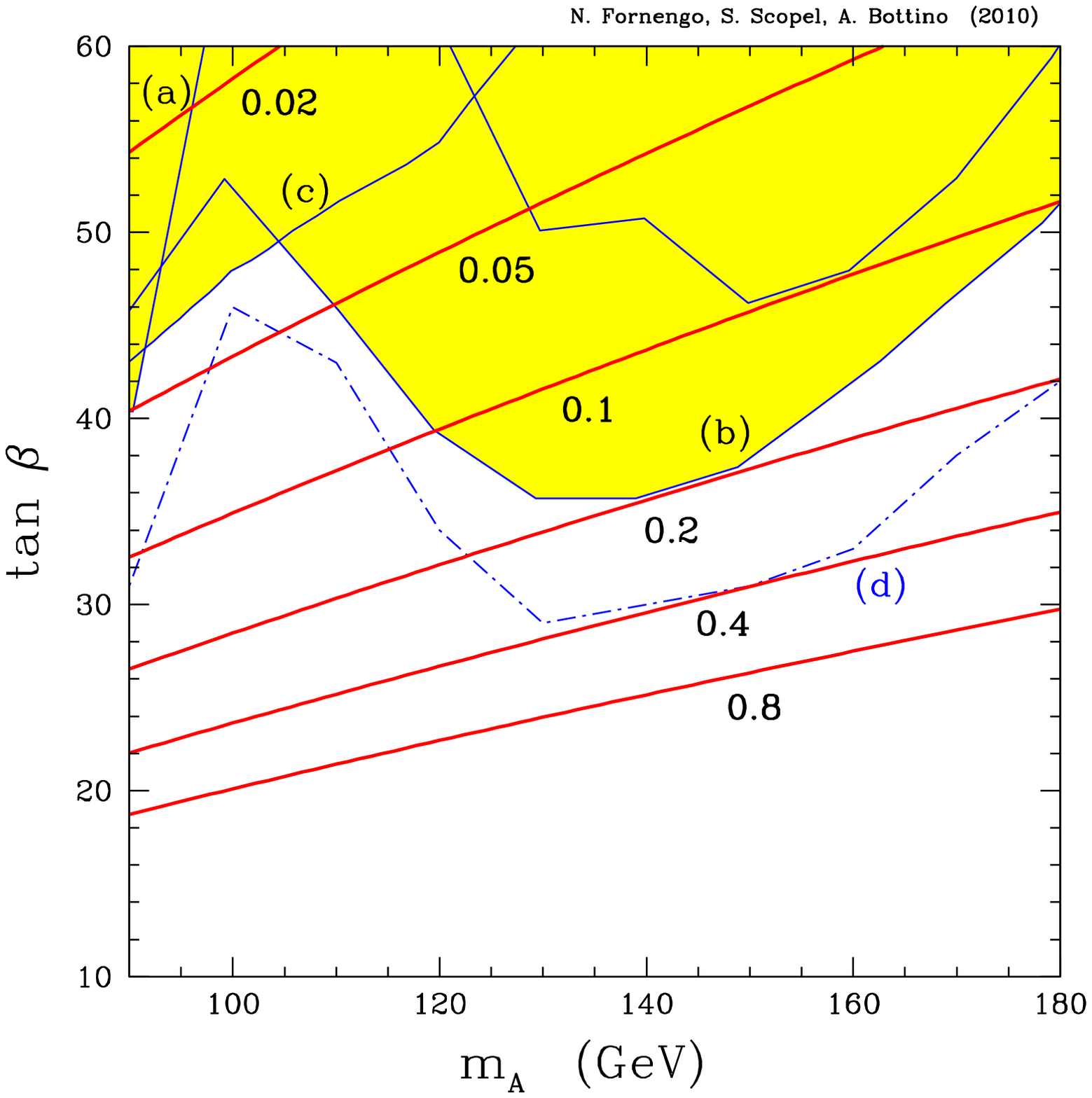}
\caption{The same as in Fig. \ref{fig:omega_tanb_higgs}, except that
curves labeled by numbers denote the upper bounds from the approximate
analytic expression of
$BR^{(6)}(B_s \rightarrow \mu^+ \mu^-)$ in Eq. (\ref{mumu4}), for
$m_{\chi^+} = 110$ GeV. The different
lines refer to values of the stop--masses splitting parameter 
$\delta = A m_t m_{\tilde q}/(m^2_{\tilde q} + m^2_t)$ in Eq. (\ref{stopratio})
as given by the corresponding number reported
close to the lines. For any given value of $\delta$, the
allowed region is below the corresponding line.}
\label{fig:bsmumu_tanb_higgs}
\end{figure}
%

The Tevatron  is expected to have  a good sensitivity to the search for Higgs neutral
bosons in the regime of small $m_A$ and large $\tan \beta$, since in this region of the supersymmetric parameters
 the couplings of the neutral Higgs bosons $\phi = h, A, H$ to the down--fermions are
 enhanced. This has prompted searches by the CDF and D0 Collaborations for the
neutral Higgs bosons which decay as $\phi \rightarrow b \bar{b}$ or
$\phi \rightarrow \tau \bar{\tau}$ (for an updated review see Ref.
\cite{bernardi}).

We report here the results of these Collaborations in terms of upper bounds on $\tan \beta$ versus $m_A$.  These bounds are displayed in Figs. 
\ref{fig:omega_tanb_higgs}, \ref{fig:scat_higgs_tanb}, \ref{fig:bsmumu_tanb_higgs} as piecewise linear paths.

The D0 Collaboration has determined upper bounds for the production rate of the
process $p \bar{p} \rightarrow \phi \rightarrow \tau^+ \tau^-$ (inclusive $\tau^+ \tau^-$
production) in Ref. \cite{abazov_tautau_2008} and for the $\tau^+ \tau^-$ production in association with a $b$ quark in Ref. \cite{abazov_tautau_2009} and then converted these bounds into
upper limits for the SUSY parameters $\tan \beta$ and $m_A$. These limits are
represented in Figs. 
\ref{fig:omega_tanb_higgs}, \ref{fig:scat_higgs_tanb}, \ref{fig:bsmumu_tanb_higgs} by the line (a) (from Ref. \cite{abazov_tautau_2008}) and the line
(c) (from Ref. \cite{abazov_tautau_2009}).

The CDF Collaboration has reported upper limits for the production rate of the
inclusive $\tau^+ \tau^-$ production in Ref. \cite{aaltonen_tautau_2009}
with ensuing upper bounds represented by the line denoted as   (b)
in Figs. 
\ref{fig:omega_tanb_higgs}, \ref{fig:scat_higgs_tanb}, \ref{fig:bsmumu_tanb_higgs}. These results supersede the stricter bounds found
by the same Collaboration in a previous analysis where $\tan \beta \lsim$ 40 at $m_A$ = 90 GeV
\cite{abulencia_tautau_2006}.

The D0 and CDF Collaborations have also presented a combined analysis of their searches for Higgs
into the inclusive $\tau^+ \tau^-$ channel  which provides upper bounds on $\tan \beta$  displaying a sharp variation at small values of $m_A$:
$\tan \beta \lsim 30-31$ at $m_A$ = 90 GeV and $\tan \beta \lsim 44-46$ at $m_A$ = 100 GeV \cite{tevnp} (see line denoted as (d) in Figs. 
\ref{fig:omega_tanb_higgs}, \ref{fig:scat_higgs_tanb}, \ref{fig:bsmumu_tanb_higgs}).

 It is worth remarking that the derivation of the bounds on the SUSY parameters
from the experimental data  require the use of a specific supersymmetric model. The one employed in Refs. \cite{abazov_tautau_2008,abazov_tautau_2009,aaltonen_tautau_2009,abulencia_tautau_2006,tevnp}
is different from the LNM;  in the scenario {$\mathcal{A}$} of our model, because of the typical small values of the parameters $\mu$ and $m_A$, the bounds on   $\tan \beta$ and $m_A$ might be more relaxed (see arguments in Sect. 3.2.1 of Ref. \cite{carena_2005}).
Notice also that the results of Ref. \cite{tevnp} are still presented as
 an (unpublished) preliminary report. Thus in our analysis we only employ the bounds of
 Refs. \cite{abazov_tautau_2008,abazov_tautau_2009,aaltonen_tautau_2009}, which taken together
 disallow the  region depicted in yellow in Figs. 
\ref{fig:omega_tanb_higgs}, \ref{fig:scat_higgs_tanb}, \ref{fig:bsmumu_tanb_higgs}.

In Fig. \ref{fig:omega_tanb_higgs} we display (as continuous curves in black) the lines where $\Omega_{\chi} h^2$, calculated with Eq. (\ref{eq:omega0}), is
equal to $(\Omega_{CDM} h^2)_{\rm max} = 0.12$ at the fixed value of $m_{\chi}$ indicated
(in units of GeV) along each curve; for the other parameters the following values are used:
 $x_f/g_{\star}(x_f)^{1/2}$ = 2.63, $a_1^2 a_3^2 = 0.12$, $\epsilon_b = -0.08$. For a given value of $m_{\chi}$ (masses from 6 GeV to 20 GeV are considered here) the region below the relevant curve is disallowed by the cosmological upper bound on $\Omega_{CDM} h^2$. By comparing the continuous (black) curves  with the Tevatron limits, one sees what is the impact
 of these limits over the allowed range for the neutralino masses. In particular,
 one notices that the (yellow) forbidden region in compatible with neutralino masses down to 7 GeV. Should one include the upper bounds of Ref. \cite{tevnp}, the lower limit on $m_{\chi}$ would be increased only very slightly to the value of about 7.5 GeV.
In Fig. \ref{fig:scat_higgs_tanb} we instead display the features of the scatter plot for the light neutralino population of LNM--$\cal A$, in the plane $m_A - \tan \beta$, once the bounds previously discussed are applied. The points denoted by red circles
refer to configurations with neutralino masses lighter than 10 GeV.

To complete the analysis of the previous Section on the bounds coming from the
$B_s \rightarrow \mu^+ + \mu^-$ decay, when combined with the Higgs searches at
the Tevatron, in Fig. \ref{fig:bsmumu_tanb_higgs} we show the upper bounds
obtained by using the approximate expression of Eq. (\ref{mumu4}) for 
$BR^{(6)}(B_s \rightarrow \mu^+ \mu^-)$. The solid lines refer to the
bounds obtained for fixed values of the stop--masses splitting parameter
$\delta = |A| m_t m_{\tilde q}/(m^2_{\tilde q} + m^2_t)$, from
$\delta = 0.02$ to $\delta=0.8$. For any value of $\delta$, the allowed region is 
below the corresponding curve. We notice that
$BR^{(6)}(B_s \rightarrow \mu^+ \mu^-)$  does not set significant bounds as long
as $\delta$ is sufficiently small, which in turn occurs for small values of $|A|$. In our scan for Scenario $\cal A$, the values
of $\delta$ naturally range between 0.01 to 0.6, for configurations
with light neutralinos.

\subsection{Search for charged Higgs bosons in top quark decay at the Tevatron}
\label{searchchargedH}

%
\begin{figure}[t]
\includegraphics[width=1.1\columnwidth,clip=true,bb=18 60 592 520]{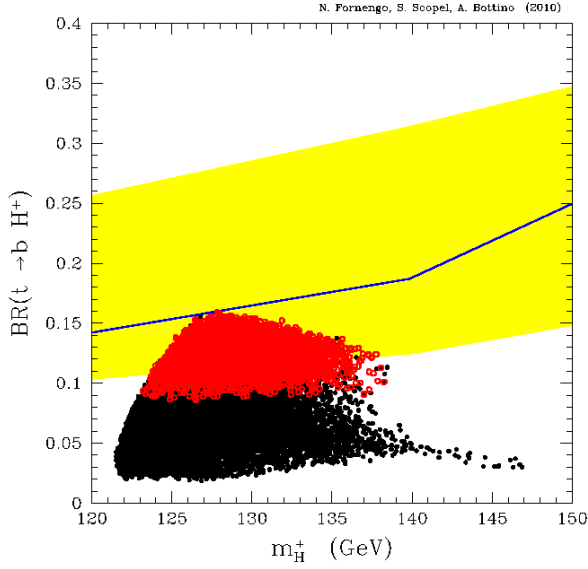}
\caption{Scatter plot of the branching ratio $BR(t \rightarrow b H^+)$ 
as a function of the charged Higgs mass $m_{H^+}$, in the LNM--$\cal A$ scan.
The solid line and the yellow band represent the experimental upper bound and
its quoted uncertainty \cite{abazov_top}. Black points stand for $m_\chi > 10$ GeV, while the red circles for
$m_\chi \leq 10$ GeV.}
\label{fig:scat_higgsc_toptohiggs}
\end{figure}
\begin{figure}[t]
\includegraphics[width=1.1\columnwidth,clip=true,bb=18 60 592 520]{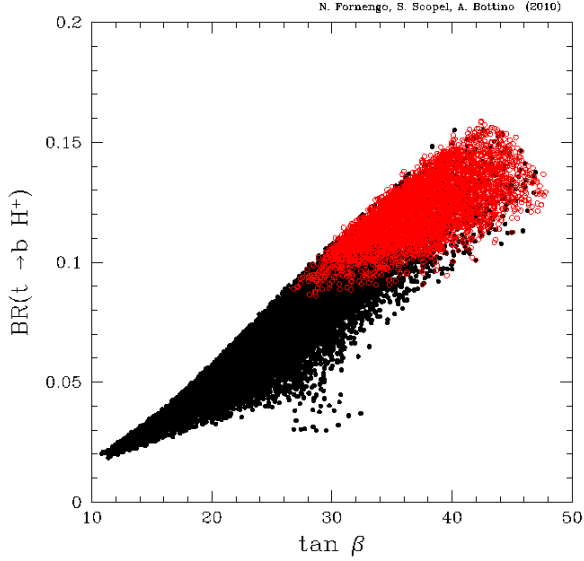}
\caption{Scatter plot of the branching ratio $BR(t \rightarrow b H^+)$ 
as a function of $\tan\beta$, in the LNM--$\cal A$ scan.
Black points stand for $m_\chi > 10$ GeV, while the red circles for
$m_\chi \leq 10$ GeV.}
\label{fig:scat_tanb_toptohiggs}
\end{figure}
\begin{figure}[t]
\includegraphics[width=1.1\columnwidth,clip=true,bb=18 60 592 520]{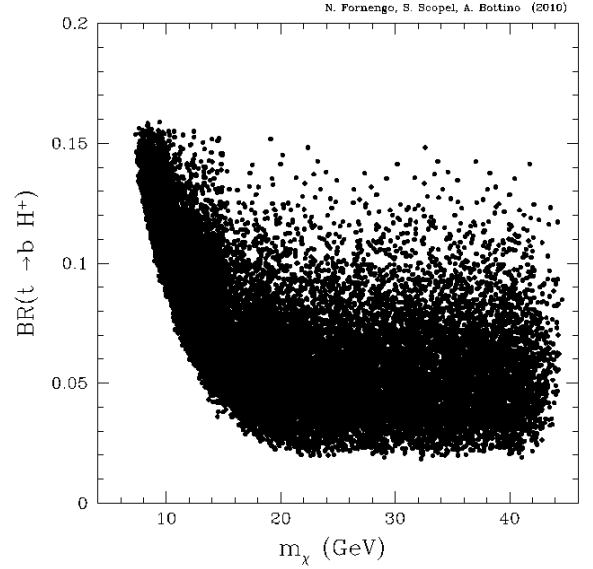}
\caption{Scatter plot of the branching ratio $BR(t \rightarrow b H^+)$ 
as a function of $m_\chi$, in the LNM--$\cal A$ scan.}
\label{fig:scat_mchi_toptohiggs}
\end{figure}
%

Supersymmetric models which contain light neutralinos automatically involve also light charged Higgs bosons $H^{\pm}$, since, at tree--level, the following relation holds $m^2_{H^{\pm}} = m^2_A + m^2_W$. This would make the decay $t \rightarrow b + H^+$ \cite{top_intob} possible in our LNM.

A search for the decay $t \rightarrow b + H^+$, conducted at the Tevatron, led the CDF Collaboration \cite{abazov_top} to establish an upper bound on $\tan \beta$ which is a monotonically increasing
function of $m_{H^{\pm}}$.
In particular, at $m_{H^{\pm}}$ = 120 GeV ({\it i.e.} at $m_A \simeq$ 90 GeV)
this constraint corresponds to
$\tan \beta \lsim$ 45--50. 
The bound on the branching ratio $BR(t \rightarrow b H^+)$, with its quoted uncertainty (yellow band), is shown in Fig. \ref{fig:scat_higgsc_toptohiggs} as a solid line, together with the scatter plot of configurations of LNM--$\cal A$.
The yellow band denotes the quoted uncertainty on the bound \cite{abazov_top}. We 
see that the current bounds on the decay $t \rightarrow b + H^+$ do not impose additional constraints on LNM--$\cal A$. Figs. \ref{fig:scat_tanb_toptohiggs} and
\ref{fig:scat_mchi_toptohiggs} show
the correlation of $BR(t \rightarrow b H^+)$ with $\tan\beta$ and with $m_\chi$,
respectively.

\subsection{$B \rightarrow \tau \nu$ and $B \rightarrow D + \tau + \nu$ at Belle and BaBar}
\label{searchB}

\begin{figure}[t]
\includegraphics[width=1.1\columnwidth,clip=true,bb=18 60 592 520]{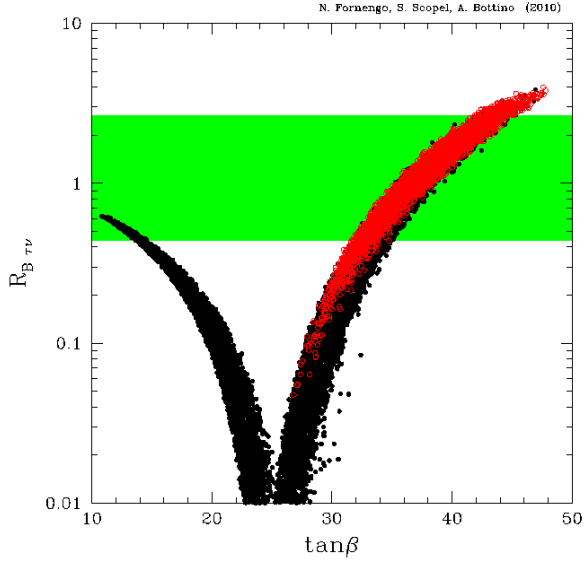}
\caption{Scatter plot of the quantity  $R_{B \tau \nu}$, calculated according
to Eqs. (\ref{Btau})--(\ref{Btau1}), as a function of $\tan \beta$ in the
LNM--$\cal A$ scan. Black points stand for $m_\chi > 10$ GeV, while the red circles for
$m_\chi \leq 10$ GeV. The green horizontal band represents the range of Eq. (\ref{Btau2}).}
\label{fig:scat_tanb_btotau}
\end{figure}
\begin{figure}[t]
\includegraphics[width=1.1\columnwidth,clip=true,bb=18 60 592 520]{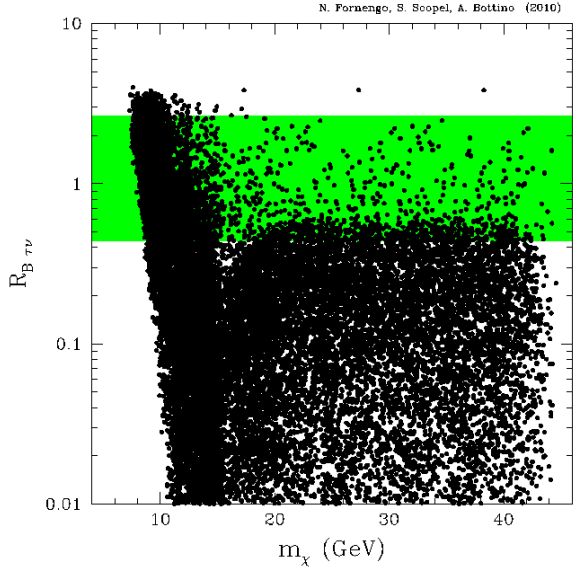}
\caption{Scatter plot of the quantity  $R_{B \tau \nu}$, calculated according
to Eqs. (\ref{Btau})--(\ref{Btau1}), as a function of $m_\chi$ in the
LNM--$\cal A$ scan. The green horizontal band represents the range of Eq. (\ref{Btau2}).
}
\label{fig:scat_mchi_btotau}
\end{figure}
\begin{figure}[t]
\includegraphics[width=1.1\columnwidth,clip=true,bb=18 60 592 520]{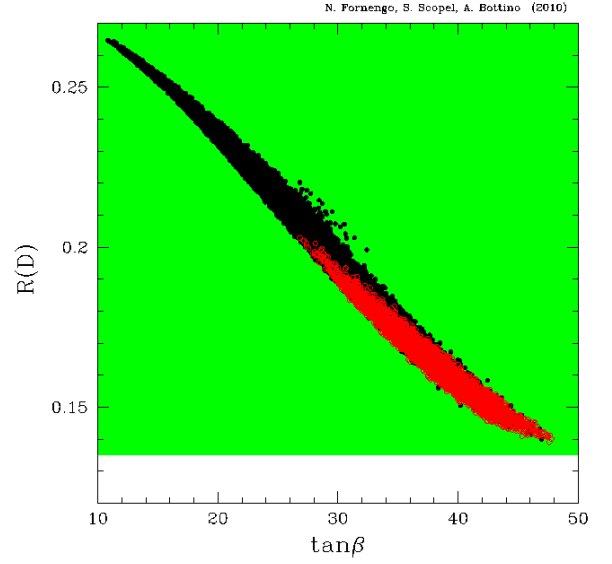}
\caption{Scatter plot of the quantity  $R(D)$, calculated according
to  Eq. (9) of Ref. \cite{kam} as a function of $\tan \beta$ in the
LNM--$\cal A$ scan. Black points stand for $m_\chi > 10$ GeV, while the red circles for
$m_\chi \leq 10$ GeV.
The green horizontal band represents the bottom part of the range of 
Eq. (\ref{BDtau}).}
\label{fig:scat_tanb_btoDtau}
\end{figure}
\begin{figure}[t]
\includegraphics[width=1.1\columnwidth,clip=true,bb=18 60 592 520]{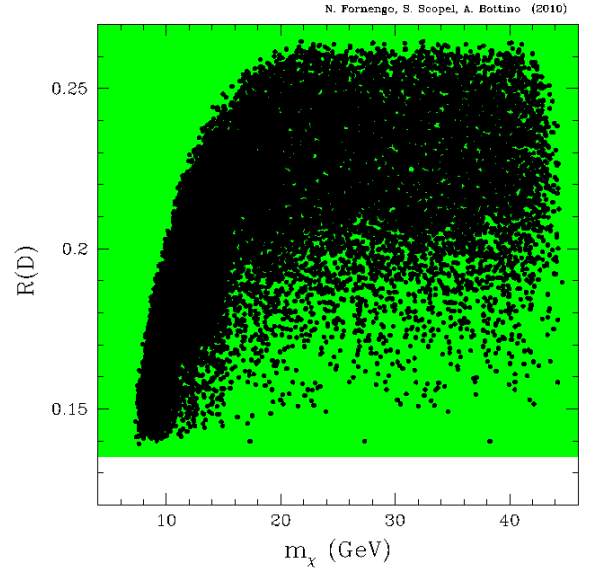}
\caption{Scatter plot of the quantity  $R(D)$, calculated according
to  Eq. (9) of Ref. \cite{kam} as a function of $m_\chi$ in the
LNM--$\cal A$ scan. 
The green horizontal band represents the bottom part of the range of 
Eq. (\ref{BDtau}).}
\label{fig:scat_mchi_btoDtau}
\end{figure}
\begin{figure}[t]
\includegraphics[width=1.1\columnwidth,clip=true,bb=18 60 592 520]{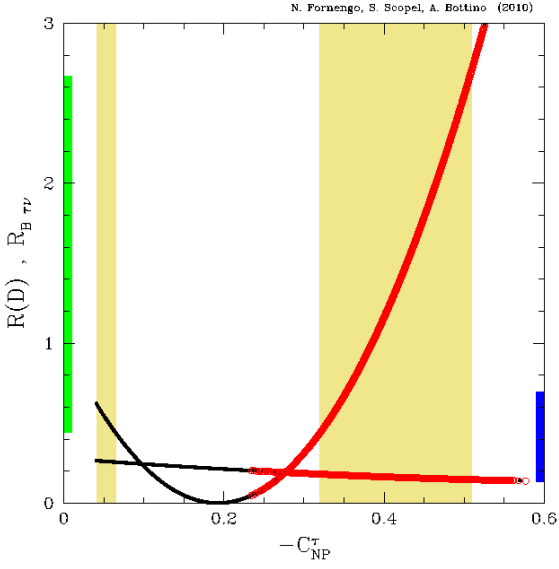}
\caption{$R_{B \tau \nu}$ and $R(D)$ as functions of $C^{\tau}_{NP}$, for configurations
of the LNM--$\cal A$ scan.
The curve with the parabolic shape and the almost straight line represent the values of
$R_{B \tau \nu}$ and  $R(D)$, respectively, for light neutralinos. The red points denote the neutralino configurations with  $m_{\chi} \leq $ 10 GeV.
The two ranges along the vertical axes denote the interval of Eq. (\ref{Btau2}) for $R_{B \tau \nu}$ (in green, on the left) and the interval of Eq. (\ref{BDtau}) for  $R(D)$
(in blue, on the right). The two vertical bands in yellow denote the ranges of $C^{\tau}_{NP}$ where the experimental intervals of $R_{B \tau \nu}$ and $R(D)$ have a common solution.}
\label{fig:scat_cp_btotau}
\end{figure}
\begin{figure}[t]
\includegraphics[width=1.1\columnwidth,clip=true,bb=18 60 592 520]{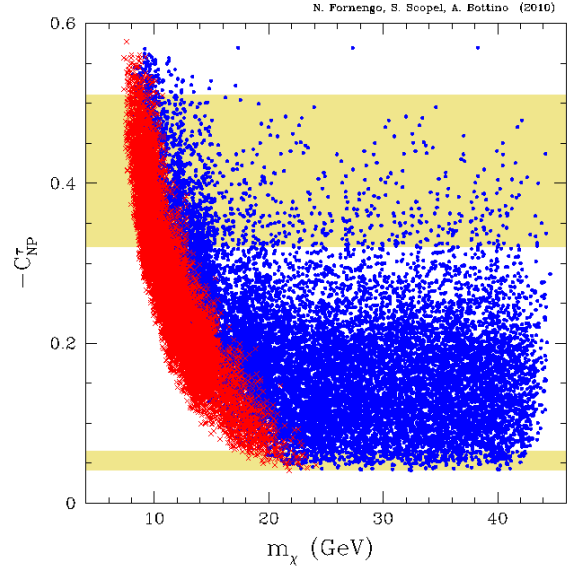}
\caption{Scatter plot of $C^{\tau}_{NP}$ as a function of the neutralino mass $m_\chi$
for the LNM--$\cal A$ scan. The two horizontal bands in yellow denote the ranges of $C^{\tau}_{NP}$ where the experimental intervals of $R_{B \tau \nu}$ and $R(D)$ have a common solution. Blue points stand for cosmologically subdominant neutralinos
({\em i.e.} $\Omega_\chi h^2 < 0.098$), while
the red crosses refer to dominant configurations
({\em i.e.} $0.098 \leq \Omega_\chi h^2 \leq 0.122$).}
\label{fig:scat_mchi_cp}
\end{figure}
\begin{figure}[t]
\includegraphics[width=1.1\columnwidth,clip=true,bb=18 60 592 520]{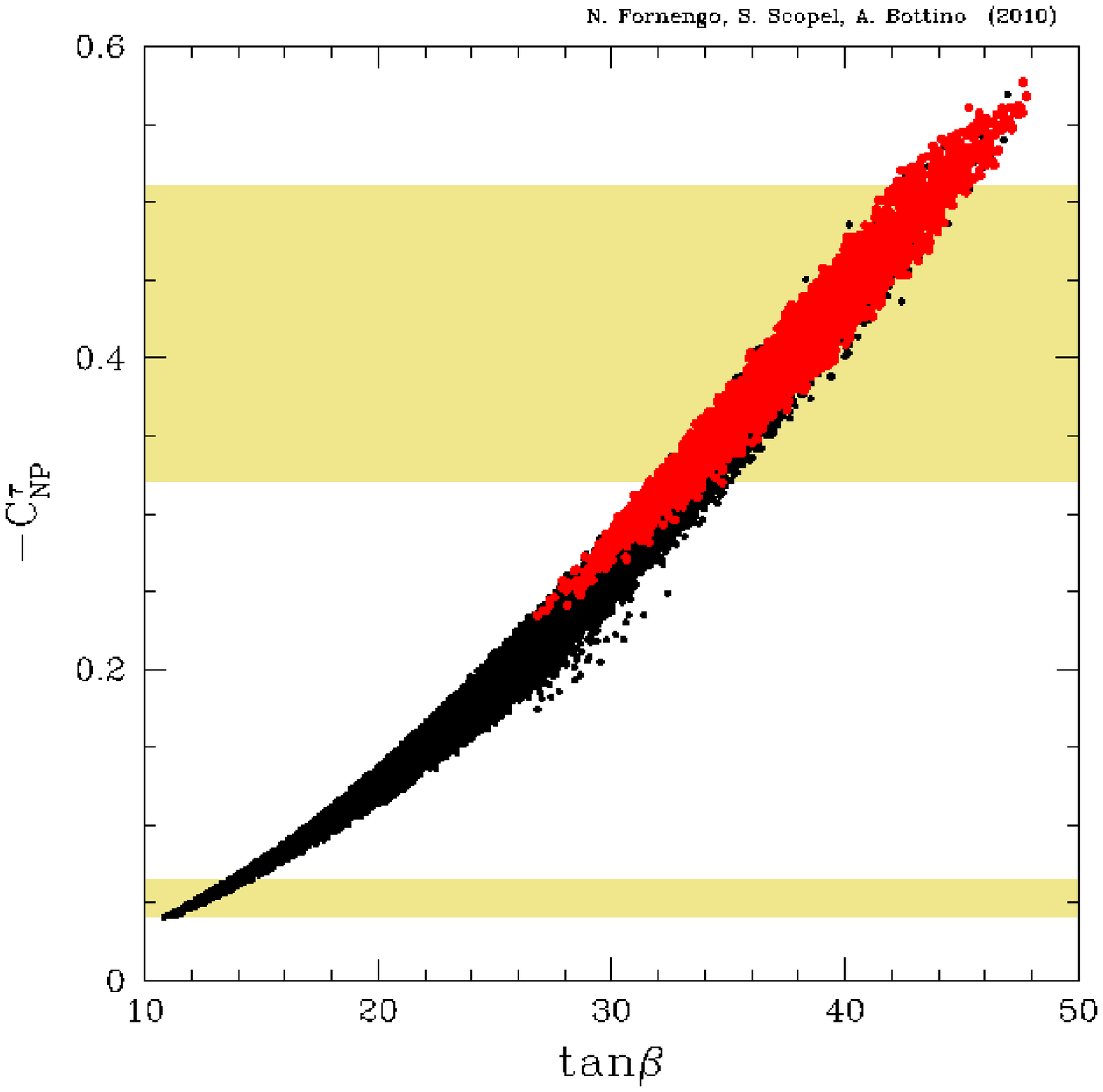}
\caption{Scatter plot of $C^{\tau}_{NP}$ as a function of $\tan\beta$
for the LNM--$\cal A$ scan. The two horizontal bands in yellow denote the ranges of $C^{\tau}_{NP}$ where the experimental intervals of $R_{B \tau \nu}$ and $R(D)$ have a common solution. Black points stand for $m_\chi > 10$ GeV, while the red circles for
$m_\chi \leq 10$ GeV.}
\label{fig:scat_tanb_cp}
\end{figure}
%

The measurements of the B-meson decays
$B \rightarrow \tau + \nu$ and $B \rightarrow D + \tau + \nu$
(and $B \rightarrow D+ l + \nu$, where $l = e,\mu$) are potentially a way to investigate
 allowed ranges for the two parameters $\tan \beta$ and $m_{H^{\pm}}$.

However, it is to be noted that at present the uncertainties  affecting the theoretical estimates as well as the experimental determinations concerning the class of B-meson decays mentioned  above imposes
 a very cautious attitude in applying {\it tout court} the entailing constraints on the SUSY parameters.
 The situation might evolve favorably in the future and thus provide either more solid constraints or hopefully a substantial indication of new physics. Thus,  we devote this section to an analysis of these processes more in view of possible prospects for the future than for an actual implementation at the present stage.

 As for the first process, a convenient quantity
 to be studied is the ratio of the total (SM contribution plus extra contributions) branching ratio
$BR_{\rm tot}(B \rightarrow \tau \nu)$ to the branching ratio due only to SM,
$BR_{SM}(B \rightarrow \tau \nu)$:
$R_{B \tau \nu} \equiv BR_{\rm tot}(B \rightarrow \tau \nu)/BR_{SM}(B \rightarrow \tau \nu)$.
If, as extra contributions, only SUSY contributions are taken, one finds
\cite{hou,akeroyd}:
\begin{equation}
R_{B \tau \nu} = \left(1 +  \frac{m^2_B}{m_b \; m_{\tau}} \; C^{\tau}_{NP}\right)^2,
\label{Btau}
\end{equation}
where:
\begin{equation}
C^{\tau}_{NP} = - \frac{m_b \; m_{\tau}}{m^2_{H^{\pm}}}
\frac{\tan^2 \beta}{1 - \epsilon_0 \; \tan \beta}.
\label{Btau1}
\end{equation}
The quantity $\epsilon_0$ is defined by:
\begin{equation}
\epsilon_0 = - \frac{2 \; \alpha_s}{3 \; \pi} \; M_3 \; \mu \; C_0(m^2_{\tilde{b}_1},m^2_{\tilde{b}_2},M^2_3),
\label{eps}
\end{equation}
where:
\begin{equation}
C_0(x,y,z) = \frac{x y \log(x/y) + y z \log(y/z) + x z \log(z/x)}{(x - y) (y - z) (z - x)}.
\label{eps1}
\end{equation}

For $R_{B \tau \nu}$ we use here the 95\% C.L. range:
\begin{equation}
0.44 \leq R_{B \tau \nu} \leq 2.67,
\label{Btau2}
\end{equation}
based on the experimental world average
$BR_{\rm exp}(B^+ \rightarrow \tau^+ \nu) = (1.72^{+ 0.43}_{-0.42}) \times 10^{-4}$ deduced in Ref. \cite{rosner}
 from the Belle \cite{belle} and  BaBar \cite{BaBar} data and
 the SM evaluation
$BR_{SM}(B^+ \rightarrow \tau^+ \nu) = (1.10 \pm 0.29) \times 10^{-4}$ \cite{alt,alt2}
(this determination for $BR_{SM}(B^+ \rightarrow \tau^+ \nu)$ is taken conservatively;
estimates by other authors \cite{rosner,martinelli} give slightly lower values).

Fig. \ref{fig:scat_tanb_btotau} and Fig. \ref{fig:scat_mchi_btotau}  display how the band of Eq. (\ref{Btau2}) compares with the population of light neutralinos in terms of $\tan \beta$ and of $m_{\chi}$,
respectively. These figures show that the present range of  $R_{B \tau \nu}$ favors large values
of $\tan \beta$ and would not have any impact on the lower bound of 7.5 GeV previously established.

As for  the  semileptonic decays $B \rightarrow D + l + \nu$, it is convenient to consider
the ratio
$R(D) \equiv BR(B \rightarrow D \tau \nu)/BR(B \rightarrow D e \nu)$,
since in this ratio many experimental systematic uncertainties cancel, either partially or completely
\cite{aubert_BD}. Also some theoretical uncertainties cancel out in this ratio \cite{kam}.

An experimental determination for $R(D)$ is provided by the  BaBar Collaboration:
$R(D) = (41.6 \pm 11.7({\rm stat}) \pm 5.2({\rm syst})) \times 10^{-2}$ \cite{aubert_BD}. A determination
by the Belle Collaboration is given in the report of Ref. \cite{iijima}:
$R(D) = (60 \pm 14({\rm stat}) \pm 8({\rm syst})) \times 10^{-2}$. However, this value 
has been obtained by using the Belle determination
$BR(B^+ \rightarrow \bar{D}^0 \tau^+ \nu) = (1.51^{+0.41}_{-0.39}({\rm stat})^{+0.24}_{-19}({\rm syst}) \pm 0.15) \times 10^{-2}$, which has
recently been superseded by the new Belle determination
$BR(B^+ \rightarrow \bar{D}^0 \tau^+ \nu) = (0.77 \pm 0.22({\rm stat}) \pm 0.12({\rm syst})) \times 10^{-2}$ \cite{belle_2010}, smaller than the previous
one by a factor of two. 
In view of the lack of an updated value for 
$R(D)$ provided by the Belle Collaboration and of the ambiguities which may rise in treating the uncertainties in the various branching ratios concurring in the determination of $R(D)$, in our discussion we only take into account the  BaBar value previously mentioned. Notice however that the new Belle value for the
branching ration of $B^+ \rightarrow \bar{D}^0 \tau^+ \nu$ approaches now considerably the  BaBar value for the branching ratio of the process $B^+ \rightarrow \bar{D}^0 \tau^+ \nu$ \cite{aubert_BD}.

From the  BaBar determination
$R(D) = (41.6 \pm 11.7({\rm stat}) \pm 5.2({\rm syst})) \times 10^{-2}$  we obtain the 95\% C.L. range:
\begin{equation}
13.5 \times 10^{-2} \leq R(D) \leq 69.7 \times 10^{-2}.
\label{BDtau}
\end{equation}

Fig. \ref{fig:scat_tanb_btoDtau} and Fig. \ref{fig:scat_mchi_btoDtau} display the scatter plot for the quantity $R(D)$
evaluated by using the Eq. (9) of Ref. \cite{kam}, versus $\tan \beta$ and $m_{\chi}$, respectively.
We see that the experimental range of $R(D)$ given in Eq. (\ref{BDtau}) is compatible with the light neutralino configurations.

In Fig. \ref{fig:scat_cp_btotau} we display  $R_{B \tau \nu}$ and $R(D)$ as functions of the quantity
$C^{\tau}_{NP}$ defined in Eq. (\ref{Btau1}). The curve with the parabolic shape represents
$R_{B \tau \nu}$ as given by Eq. (\ref{Btau}) and the almost straight line gives  $R(D)$,
calculated with  Eq. (9) of Ref. \cite{kam}. The parts colored in red pertain to neutralino
configurations with  $m_{\chi} \leq $ 10 GeV.

Notice that, should the indication of  Fig. \ref{fig:scat_cp_btotau} be taken strictly, one would deduce for the quantity $C^{\tau}_{NP}$ two ranges of compatibility:
$- 0.51 \lsim C^{\tau}_{NP} \lsim - 0.32$ and $- 0.041 \lsim C^{\tau}_{NP} \lsim - 0.065$.
From this, by using Eq. (\ref{Btau1}), one finds that the range at large $\tan\beta$
for the quantity $\tan\beta/(m_{H^{\pm}}/120 \; {\rm GeV})$ is:
$30 \lsim  \tan \beta / (m_{H^{\pm}}/120 \; {\rm GeV}) \lsim 40$.

This in turn would have  some impact on the features of the neutralino population 
 which substantially contribute to the DM abundance ({\em i.e.} for which
$(\Omega_{CDM} h^2)_{\rm min} \leq \Omega_{\chi} h^2 \leq (\Omega_{CDM} h^2)_{\rm max}$). This can be appreciated in Figs.
\ref{fig:scat_mchi_cp} and \ref{fig:scat_tanb_cp}, where $C_{NP}$ is plotted against the neutralino mass (Fig. \ref{fig:scat_mchi_cp}) or $\tan\beta$
(Fig. \ref{fig:scat_tanb_cp}). 
In Fig. \ref{fig:scat_mchi_cp}
the points denoted by red crosses refer to cosmologically dominant neutralinos, and they are compatible with the preferred ranges of $C_{NP}$ for neutralino masses below 13 GeV (and large $\tan\beta$) and in the interval $18 - 25$ GeV (and low $\tan\beta$).
In Fig. \ref{fig:scat_tanb_cp} the red circles refer to configurations with
$m_\chi \leq 10$ GeV. Notice that if we take at face value the current bounds
on the $B \rightarrow \tau$ decays, light neutralinos would actually be favored.

Nevertheless, we recall that, for the reasons mentioned at the beginning of the present Section, it seems premature to enforce these constraints rigidly at the present time (see Ref. \cite{rosner} for similar comments).

We can conclude the present Section on constraints on supersymmetric parameters from the
Tevatron collider and the B--factories with the following remarks:
i) the upper bound on the branching ratio for the decay $B_s \rightarrow \mu^+ + \mu^-$
determined at the Tevatron has a mild effect in constraining the LNM, at variance with
what occurs in SUGRA models, ii) the bounds  which are derived
 from the Higgs bosons searches at the Tevatron
 do not modify the previously mentioned lower bound
$m_{\chi} \gsim$ 7.5 GeV, iii) the measurements of the rare B--meson decays at B--factories
have still to be taken with much caution: combining  the present data on
$B \rightarrow \tau + \nu$ and $B \rightarrow D + \tau + \nu$ one derives
a range of $\tan \beta / (m_{H^{\pm}}/120 \; {\rm GeV})$ which at present might only
have some effect  on the light neutralino population for $m_{\chi} \simeq$ 15--20 GeV,
without modifying the lower bound on the neutralino mass.

We finally note that the strict bounds obtained in
 Ref. \cite{kuflik} on the WIMP--nucleon scattering cross--section are mainly due to: a) the use of
 constraints on $\tan \beta$ derived from D0--Collaboration \cite{owen} and
 CDF \cite{abulencia_tautau_2006} data which were subsequently superseded (and relaxed) by Refs. \cite{abazov_tautau_2008,aaltonen_tautau_2009}, respectively; b)  the implementation of
 bounds restrictively derived from B--meson decays data which suffer from sizable uncertainties and which have been in part superseded by the more recent measurements quoted above.

\section{Confronting results from experiments of direct detection of DM particles}
\label{sec:experiments}

%
\begin{figure}[t]
\includegraphics[width=1.1\columnwidth,clip=true,bb=18 60 592 520]{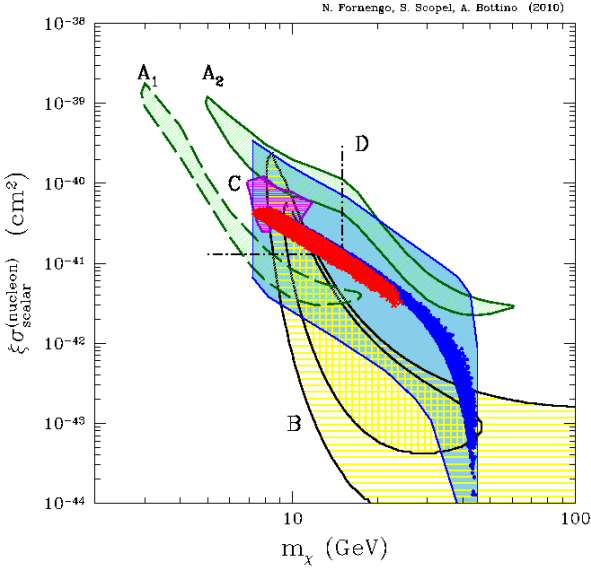}
\caption{$\xi \sigma_{\rm scalar}^{(\rm nucleon)}$ as a function of
the neutralino mass for the LNM--$\cal A$ scan and for
$g_{d,\rm ref}$ = 290 MeV (constraint from $R_{B \tau \nu}$ not included).
The (red) crosses denote configurations with a neutralino
relic abundance which matches the WMAP cold dark
matter amount (0.098 $\leq \Omega_{\chi} h^2 \leq$ 0.122),
while the (blue) dots
refer to configurations where the neutralino is subdominant
($ \Omega_{\chi} h^2 <$ 0.098).
The blue--band flag--like region denotes the
extension of the scatter plot upwards and
downwards, when the hadronic uncertainties are included.
The green shaded regions denote the DAMA/LIBRA annual modulation
regions \cite{dama_priv}; the region delimited
by the dashed (solid) line refers to the case where the channeling effect
is (is not) included.
The two regions are denoted by letters $A_1$ and $A_2$, respectively.
The yellow hatched
regions denoted by letter B display the regions (at 68\% and 85\%
C.L.) related to the two
CDMS candidates \cite{noiCDMS}.
The pink small (horizontally shaded) region denoted by letter C refers
to the CoGeNT excess of events \cite{cogent},
whereas the black straight dot--dashed lines denoted by letter D show
schematically
a region linked to the excess reported by CRESST \cite{cresst}.}
\label{fig:scat_mchi_sigmac}
\end{figure}
\begin{figure}[t]
\includegraphics[width=1.1\columnwidth,clip=true,bb=18 60 592 520]{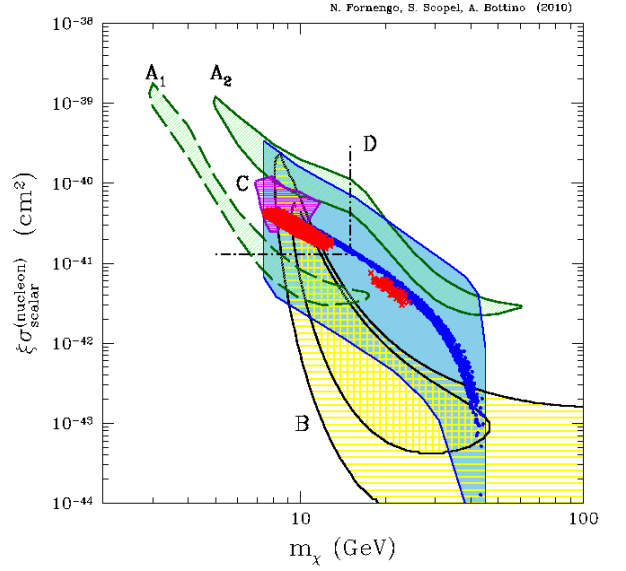}
\caption{The same as in Fig. \ref{fig:scat_mchi_sigmac} except that here the
constraint from $R_{B \tau \nu}$ is included.}
\label{fig:scat_mchi_sigmac_more}
\end{figure}
\begin{figure}[t]
\includegraphics[width=1.1\columnwidth,clip=true,bb=18 60 592 520]{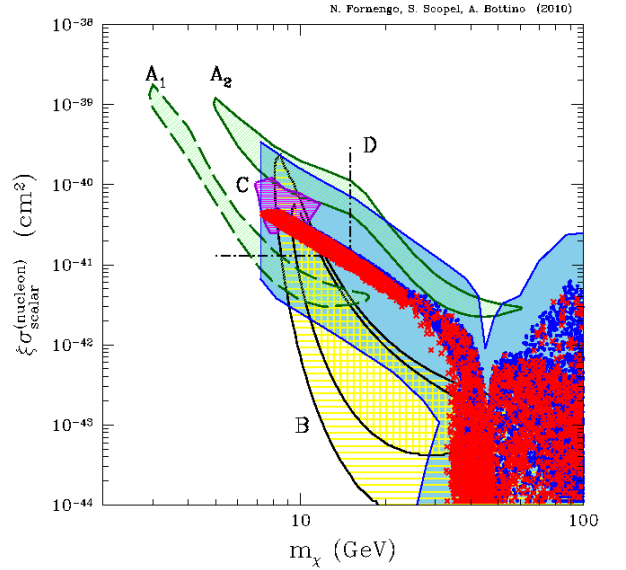}
\caption{The same as in Fig. \ref{fig:scat_mchi_sigmac} except that here the
MSSM parameter space is scanned beyond the LNM--$\cal A$
intervals, as specified in Sect. \ref{scanning}, in order to
include neutralino configurations of higher mass.}
\label{fig:scat_mchi_sigmac_full}
\end{figure}
%

As mentioned in Sect. \ref{sec:cross}, the neutralino interacts mainly
by a coherent process with the target nuclei, thus the neutralino--nucleus cross--section
is conveniently expressed in terms of the neutralino--nucleon cross--section
 $\sigma_{\rm scalar}^{(\rm nucleon)}$ and  then the
 relevant quantity to be considered is this cross--section multiplied by the rescaling factor $\xi$
 defined in Sect. \ref{sec:cross}: $\xi \sigma_{\rm scalar}^{(\rm nucleon)}$.

In Fig. \ref{fig:scat_mchi_sigmac}  we show the scatter plot representing the supersymmetric configurations
of LNM--{\cal A} and subjected to the
constraints discussed in Sects. \ref{sec:model} and \ref{sec:constraints}.
The  cross--section $\sigma_{\rm scalar}^{(\rm nucleon)}$ is calculated with its
complete expression given in Ref. \cite{bdfs2} at a fixed reference
set of values for the hadronic quantities involved in the neutralino-nucleon cross--sections
\cite{inter} (the dominant coupling $g_d$ is put at the value
$g_{d,\rm ref}$ = 290 MeV mentioned in Sect. \ref{sec:cross}).
The (red) crosses denote configurations with a neutralino
relic abundance which matches the WMAP cold dark
matter amount (0.098 $\leq \Omega_{\chi} h^2 \leq$ 0.122),
while the (blue) dots
refer to configurations where the neutralino is subdominant
($ \Omega_{\chi} h^2 <$ 0.098).
 The region covered by a (blue) slant hatching denotes the
 extension of the scatter plot upwards and
downwards, when the hadronic uncertainties extensively
discussed in Ref. \cite{uncert2} are included.

Notice that the values displayed by the scatter plot validate the
approximate expression in Eq. (\ref{bound}). They differ significantly
from the estimates given elsewhere \cite{nath,kuflik} for the reasons
discussed in the previous section.

To report in Fig. \ref{fig:scat_mchi_sigmac} also the results of
present experiments searching for direct detection of DM particles, we
have to assume a specific model for the distribution function (DF) of
the WIMPs ({\it i.e.} the neutralinos, in our case) in the galactic
halo. Among the various possible DFs \cite{bcfs}, we employ here as
reference DF the one described by the density profile of the
cored--isothermal sphere (denoted as Evans logarithmic model, or A1
model, in Ref. \cite{bcfs}) which is given by:
\begin{equation}
\rho(r) = \frac{v_0^2}{4 \pi G}\frac{3 R_c^2 + r^2}{(R_c^2 + r^2)^2}\, ,
\label{isot}
\end{equation}
where $G$ is the Newton's constant, $v_0$ is the local value of the rotational velocity and
$R_c$ is the core radius. For $R_c$ we use the value $R_c = 5$ kpc.
For the parameter $v_0$ we take the value $v_0 =  220$ km sec$^{-1}$ and
for the escape velocity
the value $v_{\rm esc} = 650$ km sec$^{-1}$. We set  $\rho = 0.34$
GeV cm$^{-3}$ for the total local DM density.

The green shaded regions denote the DAMA/LIBRA annual modulation
regions, under the hypothesis that the effect is due to a WIMP
with a coherent interaction with nuclei; the region delimited
by the solid line refers to the case where the channeling effect
is not included, the one with a dashed contour to the
case where the channeling effect is included \cite{chann}.
These regions represent the domains where the
likelihood-function values
differ more than 7.5 $\sigma$ from the null hypothesis (absence of modulation);
they are derived by the DAMA Collaboration \cite{dama_priv} from their data referring
to an exposure
of 1.17 ton x year, with an evidence for an annual modulation effect
at 8.9 $\sigma$ C.L. \cite{dama2010}.

As mentioned in the Introduction, recently a number of experimental
Collaborations have reported new data consisting of excesses of events
(over the expected backgrounds) which might represent hints for very
light DM candidates: CDMS \cite{cdms}, CoGeNT \cite{cogent}, CRESST
\cite{cresst}. Other experimental investigations (the XENON10
\cite{xenon10} and XENON100 \cite{xenon100} experiments and the CDMS
re-analyzes of previously collected data \cite{cdms_shallow}) have led
these groups to present upper bounds on $\xi \sigma_{\rm scalar}^{(\rm
  nucleon)}$ in the same range of WIMP masses (around 10 GeV).  It is
worth noting that none of these experiments is sensitive to a specific
signature of DM particles such as the annual modulation or the
directionality. Thus, their detection technique and data analysis is
based on rather intricate discrimination criteria and sizable
subtractions, which become more and more critical as the analyses of
data are extrapolated into the range of very low recoil energies. This
calls for a very cautious attitude both towards a claim of a possible
signature, in case of excesses of events over expected backgrounds, as
well as in implementing upper bounds rather fragile at the present
stage. It is worth stressing that a major critical point consists in a
reliable determination of the actual experimental efficiencies at very
low recoil energies, a problem which is the subject of much debate
\cite{collar1,xenon_reply,savage,collar2}. It is beyond the purpose of
the present paper to enter into these experimental points.  We recall
that an upper bound on $\xi \sigma_{\rm scalar}^{(\rm nucleon)}$
concerning somewhat heavier WIMPs (mass $\gsim$ 20 GeV) was also
presented by the KIMS Collaboration \cite{kims}.

In Fig. \ref{fig:scat_mchi_sigmac} we also illustrate
where the regions of the possible hints
for DM discussed in Refs. \cite{cdms,cogent,cresst} are located.
The yellow hatched
regions are related to the two CDMS candidate
events \cite{cdms}, as derived in \cite{noiCDMS} under the hypothesis that these events might be due to
DM .
Also displayed in Fig. \ref{fig:scat_mchi_sigmac} are the CoGeNT region singled out by this Collaboration as due to an excess of bulk--like events \cite{cogent} and a CRESST region which we denote with two black dot--dashed straight lines ; this region is meant to represent approximately the data reported in Ref. \cite{cresst} about  32 signals versus
a background estimate of $8.7 \pm 1.4$,  compatible with WIMPS with a mass $\lsim$ 15 GeV and a WIMP--nucleon cross--section of a few $\times 10^{-41}$ cm$^2$.

In Fig. \ref{fig:scat_mchi_sigmac_more} we display the scatter plot of the neutralino configurations, when
the constraint of Eq. (\ref{Btau2}) on $R_{B \tau \nu}$ is included. We see that
 adding this constraint would  have some impact in depriving the scatter plot of
some neutralino configurations of significant relic abundance in the range
13 GeV $\lsim m_{\chi} \lsim$ 18 GeV and for $m_{\chi} \gsim$ 25 GeV. We remind
that in Fig. \ref{fig:scat_mchi_sigmac}, \ref{fig:scat_mchi_sigmac_more}
only neutralino configurations of the special scan LNM-$\cal A$ are displayed.

Finally, in Fig. \ref{fig:scat_mchi_sigmac_full} we show the same scatter plot of 
Fig. \ref{fig:scat_mchi_sigmac}, where now
the MSSM parameter space is scanned beyond the LNM--$\cal A$
intervals, as specified in Sect. \ref{scanning}, in order to
include neutralino configurations of higher mass.

In conclusion, from the features displayed in Figs. \ref{fig:scat_mchi_sigmac},
\ref{fig:scat_mchi_sigmac_more} and \ref{fig:scat_mchi_sigmac_full} we derive that:

i) the light neutralino population agrees with the DAMA/LIBRA annual modulation data over a wide range of light neutralinos: 7--8 GeV $\lsim m_{\chi} \lsim$ 50 GeV

ii) this population is also in agreement with the data of CDMS, CoGeNT and CRESST, should these results be
significant of real DM signals; under these circumstances the range of the neutralino mass would be more
 restricted: 7--8 GeV $\lsim m_{\chi} \lsim$ (10--15) GeV.

 iii) our results contradict the argument, sometime put forward (see
 for instance Refs.~\cite{belanger,gunion}), according to which the
 results of Ref. \cite{nath}, which are obtained in the context of a
 SUGRA scenario, imply that the $BR(B_s \rightarrow \mu^+ \mu^-)$
 constraint prevents $\sigma_{\rm scalar}^{(\rm nucleon)}$ from
 reaching the size required to interpret current direct--detection
 data in terms of light neutralinos within the MSSM.

\section{Links to indirect searches for dark matter particles}

Though indirect searches for relic particles is out of the scope of
the present paper, some considerations referring to indirect effects
due to light relic neutralinos are in order here. Rather than trying
to be complete on these topics, we just recall briefly some of the
most interesting aspects.

It is known that light relic particles, through their self-annihilation
processes, can produce a flux of cosmic antiprotons in excess of
the measured one. Indeed, the experimental antiproton spectrum is fitted
well by the secondary component from cosmic rays spallation,
calculated with the set of the diffusion parameters which is derived
from the analysis of the boron--to--carbon ratio (B/C) component of
cosmic rays \cite{maurin}, with an estimated uncertainty of about 20\%.
This feature makes the cosmic antiproton flux a potential stringent
constraint for any exotic astrophysical source of primary antiprotons.

In Ref.\cite{inter} the low-mass neutralino populations extracted from the
DAMA/LIBRA data (depending of the size of the channeling effect and on
the parameters of the halo distribution function) were analyzed in
terms of the expected effects on the cosmic antiprotons. It was concluded
that many of these populations are fully compatible with the current
bounds
on cosmic antiprotons, especially for values of the local dark matter
density $\rho_0$ and local rotational velocity $v_0$ in the low side
of their physical ranges, and for values of the diffusion parameters
of the two-zone propagation model not too close to the values of their
maximal set \cite{pbar_susy}. A similar analysis in the case of the CoGeNT
data has been performed in Ref.\cite{keung} in the phenomenological
framework
of effective DM-quark interactions.

At variance with cosmic antiprotons, where primary and secondary
fluxes have very similar behaviours at low energies, and can then
hardly be disentangled from each other, measurements of cosmic
antideuterons could provide an evidence of light DM particles
\cite{dbar,dbar2008,bp}. In Ref.\cite{inter} it was shown that a
sizable number of neutralino configurations compatible with the annual
modulation data can generate signals accessible to antideuteron
searches planned for the next years.

Also the possibility of investigating light WIMPs at neutrino telescopes
has been the subject of specific investigations
\cite{hpzk,fkls,knsz,kls,nbfs}.
In Ref.\cite{nbfs} a detailed analysis of the neutrino-induced muon signal
coming from light-neutralino pair-annihilations inside the Sun and the
Earth is performed and it is shown that, under favorable conditions, a
combination of
the WIMP direct detection data and the measurements at neutrino
telescopes with
a low threshold energy could help in pinning down the features of the DM
particles.

Other possible signals due to relic light neutralinos are mentioned in
Ref.\cite{inter}.

\section{Conclusions}
\label{sec:conclusions}

In this paper we have analyzed the properties of a population of light neutralinos in an effective
MSSM at the electroweak scale, already discussed in the past \cite{lowneu,inter}, in light of new measurements at the Tevatron and B-factories which could potentially provide significant constraints
in some relevant supersymmetric parameters.

Particular attention is devoted to the branching ratio of the process
$B_s \rightarrow \mu^+ + \mu^-$ whose experimental upper bound entails  rather strict constraints on
SUGRA models. In the present analysis it is shown analitically and numerically why this experimental limit has only a mild effect on our light neutralino model.

The light neutralino population, while satisfying
the cosmological upper bound on cold dark matter, entails also a neutralino--nucleon cross--section of the
correct size to interpret the current experimental results of experiments for direct detection of
dark matter particles in terms of MSSM neutralinos.
 This population,
while fitting quite well the DAMA/LIBRA annual modulation data, would also agree with the preliminary
results of CDMS, CoGeNT and CRESST, should these data, which are at present only hints or excesses of events over the expected backgrounds,  be interpreted as authentic signals of DM.
For the neutralino mass we find a lower bound of 7-8 GeV. We have also discussed in detail
by how much this lower limit would be affected, as more refined and solid constraints from the searches on Higgs bosons and rare B-decays at the Tevatron and B-factories might be derived. It is obvious that great expectations are on the outcomes that hopefully will come out from the Large Hadron Collider at CERN to bring light to properties related to supersymmetry
(in Ref. \cite{lhc1} the perspectives of searching  for light neutralino of cosmological interest at LHC are investigated).

Our results differ from some recent conclusions by other authors;
in the course of the presentation of our results we have  tried to
single out and elucidate the main points at the origin of these variances.

\acknowledgments

A.B. and N.F. acknowledge Research Grants funded jointly by Ministero
dell'Istruzione, dell'Universit\`a e della Ricerca (MIUR), by
Universit\`a di Torino and by Istituto Nazionale di Fisica Nucleare
within the {\sl Astroparticle Physics Project} (MIUR contract number: PRIN 2008NR3EBK;
INFN grant code: FA51). S.S. acknowledges
support by NRF with CQUEST grant 2005-0049049 and by the Sogang
Research Grant 2010. N.F. acknowledges support of the spanish MICINN 
Consolider Ingenio 2010 Programme under grant MULTIDARK CSD2009- 00064.


\medskip


\begin{thebibliography}{99}

\bibitem{cdms} Z. Ahmed {\it et al.} (CDMS Collaboration), arXiv:0912.3592 and Science {327}, 1619 (2010).

\bibitem{cogent} C.E. Aalseth {\em et.~al.} (CoGeNT Collaboration), arXiv:1002.4703.

\bibitem{cresst} W. Seidel, talk given at IDM10, July 2010, Montpellier, France.

\bibitem{dama2010} R.~Bernabei {\em et al.}, Eur. Phys. J. C {\bf 67}, 39 (2010) [arXiv:1002.1028 [astro-ph.GA]]
and earlier references quoted
therein.

\bibitem{lowneu}
A. Bottino, N. Fornengo and S. Scopel, Phys. Rev. D {\bf 67}, 063519  (2003)
 [arXiv:hep-ph/0212379];
A. Bottino, F. Donato, N. Fornengo and S. Scopel, Phys. Rev. D {\bf 68}, 043506 (2003)
 [arXiv:hep-ph/0304080].

\bibitem{dama2004} R.~Bernabei {\em et al.}, Riv. Nuovo Cimento {\bf 26N1}, 1 (2003) [arXiv:astro-ph/0307403].

\bibitem{bdfs2004} A. Bottino, F. Donato, N. Fornengo and S. Scopel, Phys. Rev. D {\bf 69}, 037302 (2004) [arXiv:hep-ph/0307303].

\bibitem{zoom} A. Bottino, F. Donato, N. Fornengo and S. Scopel, Phys. Rev. D {\bf 77}, 015002 (2008) [arXiv:0710.0553[hep-ph]].

\bibitem{inter} A. Bottino, F. Donato, N. Fornengo and S. Scopel, Phys. Rev. D {\bf 78}, 083520 (2008) [arXiv:0806.4099[hep-ph]].

\bibitem{dama2008/1} R.~Bernabei {\em et al.},   Eur. Phys. J. {\bf C53}, 205 (2008) [arXiv:0710.0288 [astro-ph]].


\bibitem{dama2008/2} R.~Bernabei {\em et al.}, Eur. Phys. J. {\bf C56}, 333 (2008) [arXiv:0804.2741[astro-ph]].

\bibitem{noiCDMS} A. Bottino, F. Donato, N. Fornengo and S. Scopel,
  Phys.\ Rev.\ D {\bf 81}, 107302 (2010), [arXiv:0912.4025[hep-ph]].

\bibitem{kopp}  J. Kopp, T. Schwetz and J. Zupan, JCAP {\bf 1002}, 014 (2010) [arXiv:0912.4264 [hep-ph]].

\bibitem{fitz} A. L. Fitzpatrick, D. Hooper and K.M. Zurek, arXiv:1003.0014 [hep-ph].

\bibitem{nath} D. Feldman, Z. Liu and P. Nath, arXiv:1003.0437 [hep-ph].


\bibitem{kuflik} E.~Kuflik, A.~Pierce and K.~M.~Zurek,
  Phys.\ Rev.\  D {\bf 81}, 111701 (2010)
  [arXiv:1003.0682 [hep-ph]].

\bibitem{barger} V. Barger, M. McCaskey and G. Shaughnessy, Phys.Rev.D {\bf 82}, 035019 (2010) [arXiv:1005.3328 [hep-ph]].

\bibitem{savage} C. Savage, G. Gelmini, P. Gondolo and K. Freese, arXiv:1006.0972 [astro-ph.CO].

\bibitem{collarhooper} D. Hooper, J.I. Collar, J. Hall and D. McKinsey, arXiv:1007.1005 [hep-ph].

\bibitem{das} D. Das and U. Ellwanger, arXiv:1007.1151 [hep-ph].

\bibitem{belanger} G. Belanger {\em et.~al.}, arXiv:1008.0580 [hep-ph].

\bibitem{belikov} A.V. Belikov, J.F. Gunion, D. Hooper and T.M.P. Tait, arXiv:1009.0549 [hep-ph].

\bibitem{gunion} J.F. Gunion, A.V. Belikov,  and D. Hooper, arXiv:1009.2555 [hep-ph].

\bibitem{liu}
  P.~Draper, T.~Liu, C.~E.~M.~Wagner, L.~T.~M.~Wang and H.~Zhang,
  arXiv:1009.3963 [hep-ph].

\bibitem{chann} As for the channeling effect, it is worth mentioning that, though new investigations
have been pursued recently in connection with the competing blocking effect \cite{borzognia}, its role has still to be fully understood. The exact way
of modeling the channeling effect being still to be clarified,
a cautious attitude, which we adopted in the previous analyses and we still
adopt here,  is to consider
as physically possible the situations bracketed by the case where channeling is negligible
(no-channeling, for short) and the case where the channeling is modeled as done
in  Ref. \cite{dama2008/1}.

\bibitem{borzognia} N. Borzognia, G.B. Gelmini and P. Gondolo, arXiv:1006.3110 [astro-ph.CO] and
arXiv:1008.3676 [astro-ph.CO].


\bibitem{albornoz} D. Albornoz Vasquez, G. Belanger, C. Boehm, A. Pukhov and Silk, arXiv:1009.4380 [hep-ph].

\bibitem{LEPb} A. Colaleo (ALEPH Collaboration), talk at
SUSY'01, June 11-17, 2001, Dubna, Russia; J. Abdallah et al.
(DELPHI Collaboration), DELPHI 2001-085 CONF 513, June 2001;
LEP Higgs Working Group for Higgs boson searches, arXiv:hep-ex/0107029;
LEP2 Joint SUSY Working Group, {\tt http://lepsusy.web.cern.ch/lepsusy/}.

\bibitem{aleph05}  The ALEPH Collaboration, the DELPHI Collaboration, the L3 Collaboration,
the OPAL Collaboration, the SLD Collaboration,
the LEP Electroweak Working Group, the SLD electroweak, heavy flavour groups,
Phys.Rept. {\bf 427}:257,2006 	[arXiv:hep-ex/0509008].


\bibitem{pdg} K. Nakamura {\it et al.} (Particle Data Group), J. Physics G {\bf 37}, 075021 (2010).

\bibitem{bsgamma} E. Barberio {\em et al.} (HFAG), arXiv:hep-ex/0603003.


\bibitem{bsgamma_theorySUSY}
M. Ciuchini, G. Degrassi, P. Gambino and G.F. Giudice, Nucl. Phys. B {\bf 534}, 3 (1998) [arXiv:hep-ph/9806308].

\bibitem{bsgamma_theorySM}
M. Misiak {\em et al.}, Phys. Rev. Lett. {\bf 98}, 022002 (2007) [arXiv:hep-ph/0609232].



\bibitem{bennet} G.W. Bennet {\em et al.} (Muon g-2 Collaboration), Phys. Rev. D {\bf} 73, 072003 (2006)
[arXiv:hep-ex/0602035].

\bibitem{davier} M. Davier, A. Hoecker, B. Malaescu, C.Z. Yuan and Z. Zhang, arXiv:0908.4300 [hep-ph].

\bibitem{moroi} T. Moroi, Phys.Rev. D53 (1996) 6565-6575; Erratum-ibid. D56 (1997) 4424 [arXiv:hep-ph/9512396].

\bibitem{abazov_tautau_2008} V.M. Abazov {\em et al.} (D0 Collaboration), Phys.Rev.Lett.101, 071804 (2008)  [arXiv:0805.2491 [hep-ex]].


\bibitem{abazov_tautau_2009} V.M. Abazov {\em et al.} (D0 Collaboration), Phys.Rev.Lett.104,151801 (2010)  [arXiv:0912.0968 [hep-ex]].

\bibitem{aaltonen_tautau_2009} T. Aaltonen {\em et al.} (CDF Collaboration), Phys.Rev.Lett.103:201801,2009 [arXiv:0906.1014 [hep-ex]]


\bibitem{cdf_mumu} T. Aaltonen {\em et al.} (CDF Collaboration),
Phys. Rev. Lett.  {\bf 100}, 101802 (2008), [arXiv:0712.1708 [hep-ex]].


\bibitem{buras_delta} A.J. Buras, P.H. Chankowski, J. Rosiek and L. Slawianowska, Phys. Lett. B {\bf 546}, 96 (2002)
[arXiv:hep-ph/0207241].


\bibitem{isidori} 
  G.~Isidori and P.~Paradisi,
  Phys.\ Lett.\  B {\bf 639}, 499 (2006)
  [arXiv:hep-ph/0605012].

\bibitem{lhc1}
A.~Bottino, N.~Fornengo, G.~Polesello, and S.~Scopel,  Phys. Rev. D {\bf 77},
 115026 (2008) [arXiv:0801.3334[hep-ph]].

\bibitem{higgs}  A.~Bottino, N.~Fornengo and S.~Scopel, Nucl. Phys. B {\bf 608},
461 (2001) [arXiv:hep-ph/0012377].

\bibitem{wmap} J. Dunkley {\em et al.} (WMAP Collaboration), Astrophys.J.Suppl. {\bf 180}, 306 (2009)
[arXiv:0803.0586 [astro-ph]].

\bibitem{giapponese} H. Komatsu, Phys. Lett. B {\bf 177}, 201 (1986).

\bibitem{barbieri} R. Barbieri, G. Gamberini, G.F. Giudice and
 G. Ridolfi, Phys. Lett. B {\bf 195}, 500 (1987).


\bibitem{dreiner}Compatibility of very light neutralinos with the
  invisible width for $Z\rightarrow \chi + \chi$ (and other collider
  and low-energy bounds) was also discussed in H.~K.~Dreiner,
  S.~Heinemeyer, O.~Kittel, U.~Langenfeld, A.~M.~Weber and
  G.~Weiglein,
  Eur.\ Phys.\ J.\  C {\bf 62}, 547 (2009)
  [arXiv:0901.3485 [hep-ph]].


\bibitem{hooper} D. Hooper and T. Plehn, Phys. Lett. {\bf B562} (2003) 18 	[arXiv:hep-ph/0212226].

\bibitem{boudjema}  G. Belanger, F. Boudjema, A. Pukhov and S. Rosier-Lees, arXiv:hep-ph/0212227.

 \bibitem{uncert2} A. Bottino, F. Donato, N. Fornengo and S. Scopel,
 Astrop. Phys. {\bf 13}, 215 (2000)  [arXiv:hep-ph/9909228];
   Astrop. Phys. {\bf 18}, 205 (2002) [arXiv:hep-ph/0111229].

\bibitem{bdfs2}
A. Bottino, F. Donato, N. Fornengo and S. Scopel,
   Phys. Rev. {\bf D59}, 095003 (1999) [arXiv:hep-ph/9808456].

\bibitem{efo} J.R. Ellis, A. Ferstel and K.A. Olive, Phys. Lett. B {\bf 481}, 304 (2000) [arXiv:hep-ph/0001005]

\bibitem{rescaling} A. Bottino, V. de Alfaro, N. Fornengo, G. Mignola and S. Scopel,
Astroparticle Physics {\bf 2}, 77 (1994) [arXiv:hep-ph/9309219].

\bibitem{gaisser} T.K. Gaisser, G. Steigman and S. Tilav, Phys. Rev. D {\bf 34}, 2206 (1986).

\bibitem{bobeth} C. Bobeth, T. Ewerth, F. Kruger and J. Urban, Phys. Rev. D {\bf 64}, 074014 (2001) [arXiv:hep-ph/0104284].

\bibitem{arnowitt} R. Arnowitt, B. Dutta, T. Kamon and M. Tanaka, Phys.Lett.B {\bf 538}, 121 (2002) [arXiv:hep-ph/0203069].

\bibitem{buras} A.J. Buras, P.H. Chankowski, J. Rosiek and L. Slawianowska, Nucl.Phys. B {\bf 659}, 3 (2003)
[arxiv:hep-ph/0210145].

\bibitem{eos} J.R. Ellis, K.A. Olive and V.C. Spanos, Phys.Lett.B {\bf 624}, 47 (2005) [arXiv:hep-ph/0504196].

\bibitem{cdf_note} http://www-cdf.fnal.gov

\bibitem{bernardi} G. Bernardi, M. Carena and T. Junk, in K. Nakamura {\it et al.} (Particle Data Group), J. Physics G {\bf 37}, 075021 (2010).

\bibitem{abulencia_tautau_2006} A. Abulencia {\em et al.} (CDF Collaboration), Phys.Rev.Lett.96, 011802 (2006)  [arXiv:hep-ex/0508051].

\bibitem{tevnp} The CDF Collaboration, The D0 Collaboration, The Tevatron New Physics Higgs Working Group (TEVNPHWG), arXiv:1003.3363 [hep-ex]

\bibitem{carena_2005} M. Carena, S. Heinemeyer, C.E.M. Wagner and G. Weiglein, Eur.Phys.J.C {\bf 45}, 797 (2006)  [arXiv:hep-ph/0511023].

\bibitem{top_intob} M. Carena, D. Garcia, U. Nierste and C.E.M. Wagner, Nucl. Phys. B {\bf 577}, 88 (2000) [arXiv:hep-ph/9912516].

\bibitem{abazov_top} V. Abazov {\em et al.} (D0 Collaboration), Phys.Lett.B {\bf 682}, 278 (2009) 
[arXiv:0908.1811 [hep-ex]].

\bibitem{hou} W.S. Hou, Phys. Rev. D {\bf 48}, 2342 (1993).

\bibitem{akeroyd} A.G. Akeroyd and F. Mahmoudi, JHEP {\bf 0904}, 121 (2009) [arXiv:0902.2393 [hep-ph]] .

\bibitem{rosner} J.L. Rosner and S. Stone, [arXiv:1002.1655 [hep-ph]].

\bibitem{belle} K. Ikado {\em et al.} (Belle Collaboration), Phys. Rev. Lett. {\bf 97}, 251802 (2006)
[arXiv:hep-ex/0604018];
I. Adachi {\em et al.} (Belle Collaboration), arXiv:0809.3834.

\bibitem{BaBar} B. Aubert {\em et al.} (BaBar Collaboration), Phys. Rev. D {\bf 77}, 011107R (2008)
[arXiv:0708.2260 [hep-ex]] and [arXiv:0912.2453].

\bibitem{alt} W. Altmannshofer and D.M. Straub, arXiv:1004.1993 [hep-ph].

\bibitem{alt2} W. Altmannshofer, A.J. Buras, S. Gori, P. Paradisi and D.M. Straub, Nucl Phys. B {\bf 38}, 17 (2010) [arXiv:0909.1333 [hep-ph]].

\bibitem{martinelli} M. Bona {\em et al.}, arXiv:0908.3470 [hep-ph].

\bibitem{aubert_BD} B. Aubert {\em et al.} (BaBar Collaboration), Phys.Rev.D
{\bf 79}, 092002 (2009) [arXiv:090-2.2660 [hep-ex]].

\bibitem{kam} J.F. Kamenik and F. Mescia, Phys. Rev. D {\bf 78}, 014003 (2008) [arXiv:0802.3790 [hep-ph]].

\bibitem{iijima} T. Iijima, talk given at the Lepton-photon 2009 Symposium, August 2009,
Hamburg [http://lp09.desy.de].

\bibitem{belle_2010} A. Bozek {\em et al.} (Belle Collaboration), arXiv:1005.2302 [hep-ex].

\bibitem{owen}  M. Owen (D0 Collaboration), arXiv:0705.2329 [hep-ph].

\bibitem{bcfs} P. Belli, R. Cerulli, N. Fornengo and S. Scopel,
Phys. Rev. D {\bf 66}, 043503 (2002) [arXiv:hep-ph/0203242].

\bibitem{dama_priv} DAMA Collaboration (private communication).



\bibitem{xenon10} J. Angle {\em et al.} (XENON Collaboration),
  Phys. Rev. Lett. {\bf 100}, 021303 (2008) [arXiv:0706.0039
  [astro-ph]]; J. Angle {\em et al.} (XENON10 Collaboration),
  Phys. Rev. D {\bf 80}, 115005 (2009) [arXiv:0910.3698
  [astro-ph.CO]].

\bibitem{xenon100} E. Aprile et al. (XENON100 Collaborations), arXiv:1005.0380 [astro-ph.CO].

\bibitem{cdms_shallow} D.S. Akerib {\em et al.} (CDMS Collaboration), arXiv:1010.4290 [astro-ph.CO]
and arXiv:1011.2482 [astro-ph.CO].

\bibitem{collar1} J.I. Collar and D.N. McKinsey, arXiv:1005.0838 [astro-ph.CO].

\bibitem{xenon_reply} The XENON100 Collaboration, arXiv:1005.2615 [astro-ph.CO].

\bibitem{collar2} J.I. Collar, arXiv:1006.2031 [astro-ph.CO].

\bibitem{kims} H.S. Lee {\em et al.} (KIMS Collaboration), Phys. Rev. Lett. {\bf 99}, 091301 (2007)
 [arXiv:0704.0423v2 [astro-ph]].

\bibitem{maurin} D. Maurin, F. Donato, R. Taillet and P. Salati,
Astroph. J. {\bf 555}, 585 (2001) [arXiv:astro-ph/0101231].

\bibitem{pbar_susy} F. Donato, N. Fornengo, D. Maurin, P. Salati and
  R. Taillet, Phys. Rev. D {\bf 69}, 063501 (2004)
  [arXiv:astro-ph/0306207].

\bibitem{keung} W.-Y. Keung, I. Low and G. Shaughnessy, arXiv:1010.1774
[hep-ph].

\bibitem{dbar}
  F.~Donato, N.~Fornengo and P.~Salati,
  Phys.\ Rev.\  D {\bf 62}, 043003 (2000)
  [arXiv:hep-ph/9904481].

\bibitem{dbar2008}
  F.~Donato, N.~Fornengo and D.~Maurin,
  Phys.\ Rev.\  D {\bf 78}, 043506 (2008)
  [arXiv:0803.2640 [hep-ph]].


\bibitem{bp}
 H.~Baer and S.~Profumo,
  JCAP {\bf 0512}, 008 (2005)
  [arXiv:astro-ph/0510722].

\bibitem{hpzk} D. Hooper, F. Petriello, K.M. Zurek and M. Kamionkowski,
Phys. Rev. D {\bf 79}, 015010 (2009) [arXiv:0808.2464 [hep-ph]].

\bibitem{fkls} J.L. Feng, J. Kumar, J. Learned and L.E. Strigari, JCAP
0901:032 (2009)
[arXiv:0808.4151 [hep-ph]].

\bibitem{knsz} J. Kopp, V. Niro, T. Schwetz and J. Zupan, Phys.Rev.D
{\bf 80},083502 (2009)
[arXiv:0907.3159 [hep-ph]].

\bibitem{kls} 
J.~Kumar, J.~G.~Learned and S.~Smith,
Phys.\ Rev.\ D {\bf 80}, 113002 (2009) [arXiv:0908.1768 [hep-ph]].

\bibitem{nbfs} V. Niro, A. Bottino, N. Fornengo and S. Scopel,
Phys.Rev.D {\bf 80},
095019 (2009) [arXiv:0909.2348 [hep-ph]]. 


\end{thebibliography}
\end{document}